\documentclass[11pt,amsmath,amssymb,aps, prb,unsortedaddress]{revtex4-1}
\usepackage{physics,graphicx,enumerate,comment,dcolumn,bbold,bm}
\usepackage{array,epsfig,wrapfig,color,soul,verbatim,mathtools,mathrsfs,textcomp,float}
\usepackage[T1]{fontenc}
\usepackage[latin9]{inputenc}
\usepackage{amsmath}
\usepackage{graphicx}
\usepackage{esint}
\usepackage{color}
\usepackage{dcolumn}
\usepackage[english]{babel}
\usepackage[mathscr]{euscript}
\excludecomment{toexclude}
\usepackage{changepage} 
\usepackage[framed]{matlab-prettifier}

\newcommand{\unitmat}{\hbox{{\sf 1}\kern-.22em{\sf I}}}

\begin{document}

\title{Path integral Monte Carlo method for the quantum anharmonic oscillator}

\author{Shikhar Mittal}
\address{The Blackett Laboratory, Imperial College London, London SW7 2AZ, United Kingdom}

\author{Marise J. E. Westbroek}
\affiliation{The Blackett Laboratory, Imperial College London, London SW7 2AZ, United Kingdom}
\affiliation{Department of Earth Science and Engineering, Imperial College London, London SW7 2BP, United Kingdom}

\author{Peter R. King}
\affiliation{Department of Earth Science and Engineering, Imperial College London, London SW7 2BP, United Kingdom}

\author{Dimitri D. Vvedensky}
\affiliation{The Blackett Laboratory, Imperial College London, London SW7 2AZ, United Kingdom}

\begin{abstract}

The Markov chain Monte Carlo (MCMC) method is used to evaluate the imaginary-time path integral of a quantum oscillator with a potential that includes both a quadratic term and a quartic term whose coupling is varied by several orders of magnitude.  This path integral is discretized on a time lattice and calculations for the energy and probability density of the ground state and energies of the first few excited states are carried out on lattices with decreasing spacing to estimate these quantities in the continuum limit.  The variation of the quartic coupling constant produces corresponding variations in the optimum simulation parameters for the MCMC method and in the statistical uncertainty for a fixed number of paths used for measurement. The energies and probability densities are in excellent agreement with those obtained from numerical solutions of Schr\"odinger's equation.

\end{abstract}
\maketitle

\section{Introduction}
\label{sec1}

The harmonic oscillator is one of the few exactly solvable quantum mechanical systems.  Solutions for the energy eigenvalues and wave functions are obtained either by the analytic solution of the time-independent Schr\"odinger equation in terms of Hermite polynomials, or through a more abstract algebraic procedure based on raising and lowering operators \cite{merzbacher98}.  The operator method makes a direct connection to the statistics of bosons, a conceptual building block of modern physics.  The bosonic character of the quantum harmonic oscillator is used in the quantization of small-amplitude vibrations in molecules and solids, the latter leading to phonons \cite{ashcroft76}, the quantum theory of radiation\cite{fermi32,bohm89,feynman63}, quantum field theory \cite{zee10}, and as an illustration of the correspondence principle \cite{liboff84}.  Such diverse applications attest to the pivotal role of the harmonic oscillator in the conceptual and computational development of quantum physics.

Sufficiently small fluctuations in any system around a stable equilibrium point may be described in terms of decoupled harmonic oscillators (normal modes), regardless of the shape of the confining potential.  However, there are regimes where the harmonic oscillator paradigm breaks down.  For example, during the thermal expansion of solids, the transformations between solid phases, and chemical reactions, the displacements of atoms from their equilibrium positions cannot be regarded as small.  As harmonic interactions are derived by truncating the Taylor series expansion of the interatomic potential at second order \cite{ashcroft76}, we look to higher-order terms to augment the interaction potential. A basic one-dimensional model that incorporates the leading (quartic) correction to the harmonic potential has the Schr\"odinger equation,
\begin{equation}
\hat{H}\psi=\biggl(-{\hbar^2\over2m}{d^2\over dx^2}+{m\omega^2x^2\over2}+\lambda x^4\biggr)\psi=E\psi\, ,
\label{eq1}
\end{equation}
in which $m$ is the mass of the particle, $\omega=(k/m)^{1/2}$ is the natural frequency of the harmonic oscillator, where $k$ is the stiffness of the potential, and $\lambda$ is the coupling constant for the quartic term of the potential.

In the absence of an elementary analytic method for obtaining solutions of (\ref{eq1}), attention turned to various approximate calculations. In Rayleigh--Schr\"odinger perturbation theory, the ground-state energy $E_0(\lambda)$ of (\ref{eq1}) has the formal expansion
\begin{equation}
E_0(\lambda)={\textstyle{1\over2}}\hbar\omega+\hbar\omega\sum_{n=1}^\infty A_n\biggl({\lambda\hbar\over m\omega^2}\biggr)^n\, ,
\label{eq2}
\end{equation}
where the expansion coefficients $A_n$ must be determined.  Bender and Wu calculated \cite{bender69} these coefficients to 75th order and found a rapid increase in $|A_n|$, which suggests that (\ref{eq2}) diverges.  In fact, they showed that this series diverges for any $\lambda>0$ because of the singularity at $\lambda=0$ that separates the regions where (\ref{eq1}) has an infinite sequence of bound states $(\lambda>0)$ from the region where there are only metastable states that can escape to infinity $(\lambda<0)$ \cite{bender69,bender73,bender78}.  This explanation is a particular case of an argument first advanced by Dyson\cite{dyson52} for perturbation expansions in quantum electrodynamics. The divergence of (\ref{eq2}) has provided the impetus for alternative perturbation expansions \cite{halliday80}, resummation methods \cite{simon70}, and other computational schemes\cite{hioe75,banerjee78} for quantum anharmonic oscillators.

In this paper, we adopt a somewhat different approach to solving (\ref{eq1}) by evaluating the imaginary-time path integral for this system using the Markov chain Monte Carlo (MCMC) method \cite{creutz81}. Calculations are carried out on lattices with decreasing spacing to estimate the energy and probability density of the ground state and energies of low-lying excited states  in the continuum limit.  Comparisons with numerical integrations of Schr\"odinger's equation demonstrate the accuracy of our approach. These calculations are also pedagogical:~the wide range of $\lambda$-values in (\ref{eq1}) used in this study help to develop an intuition about how the potential affects parameters and convergence.

Although we are studying a specific case, the MCMC method can be applied to a quantum mechanical particle in any potential with minimal change to the basic procedure described here \cite{creutz81}.  Imaginary-time time path integrals can also be formulated for classical dynamics \cite{phythian77,peliti78,jensen81,westbroek18a}, as well as providing a bridge to quantum field theory \cite{zinn05}.

The organization of this paper is as follows.  We briefly outline the derivation of real-time and imaginary-time path integrals in Sec.~\ref{sec2}, referring to our earlier work\cite{westbroek18b,westbroek18c} for a more comprehensive discussion.  The MCMC method is summarized in Sec.~\ref{sec3}, and in Sec.~\ref{sec4} we discuss how varying the coupling constant of the quartic term in the potential energy affects the parameters in this method.  Section~\ref{sec5} presents the correlation functions that can be calculated from the imaginary-time formalism to obtain the energy eigenvalues of (\ref{eq1}).  We have calculated the energy and probability density of the ground state and the energies of the first two excited states.  We summarize our results and discuss other applications of imaginary-time path integrals in Sec.~\ref{sec5}.

\section{Real-time and imaginary-time path integrals}
\label{sec2}

\subsection{Feynman path integral}
\label{sec2.1}

The time-dependent Schr\"odinger equation,
\begin{equation}i\hbar{\partial\psi\over\partial t}=\hat{H}\psi\, ,
\end{equation}
for a Hamiltonian operator
\begin{equation}
\hat{H}={\hat{p}^2\over2m}+V(\hat{x})\, ,
\label{eq4}
\end{equation}
has the formal solution
\begin{equation}
\psi(x,t)=e^{-i\hat{H}t/\hbar}\psi(x,0)\, ,
\end{equation}
where the exponential factor is the evolution operator. The connection to Feynman's path integral is made through the matrix elements of the evolution operator between any two initial and final position eigenstates. In Dirac's bra-ket notation \cite{westbroek18c},
\begin{equation}
\langle x_{\!f},t_{\!f}|x_i,t_i\rangle=\langle x_{\!f}|e^{-i\hat{H}(t_{\!f}-t_i)/\hbar}|x_i\rangle
=\int [Dx(t)]\exp\biggl[-{i\over\hbar}\int_{t_i}^{t_{\!f}}L(x(t))\,dt\biggr]\, ,
\label{eq6}
\end{equation}
in which $L$ is the classical Lagrangian corresponding to the Hamiltonian operator (\ref{eq4}):
\begin{equation}
L(x(t))={m\over2}\biggl({dx\over dt}\biggr)^2-V(x(t))\, .
\label{eq7a}
\end{equation}
The notation $[Dx(t)]$ in (\ref{eq6}) means that the integral includes all space-time paths $(x,t)$ between $(x_i,t_i)$ and $(x_{\!f},t_{\!f})$.  The phase of each path is determined by the classical action $S=\int_{t_i}^{t_{\!f}}L(x(t))\,dt$ over that path. The path integral on the right-hand side of (\ref{eq6}) with the Lagrangian (\ref{eq7a}) was first derived by Feynman \cite{feynman48,hibbs65,wenzel}.

\subsection{Imaginary-time path integrals}
\label{sec2.2}

An alternative formulation of path integrals uses imaginary time, where $t$ is replaced by $-i\tau$ \cite{brush61}.  The imaginary-time path integral analogous to (\ref{eq6}) is
\begin{equation}
\langle x_{\!f},\tau_{\!f}|x_i,\tau_i\rangle=\langle x_{\!f}|e^{-\hat{H}(\tau_{\!f}-\tau_i)/\hbar}|x_i\rangle
=\int[Dx(\tau)]\exp\biggl[-{1\over\hbar}\int_{\tau_i}^{\tau_{\!f}}L_E(x(\tau))\,d\tau\biggr]\, ,
\label{eq8}
\end{equation}
in which the classical Euclidean ``Lagrangian'' is
\begin{equation}
L_E(x(\tau))={m\over2}\biggl({dx\over d\tau}\biggr)^2+V(x(\tau))\, .
\label{eq9}
\end{equation}
Just as for real times, the integral in (\ref{eq8}) is over all paths between the initial point $x_i$ and the final point $x_{\!f}$. 

The connection to quantum statistical mechanics is obtained by setting $x_{\!f}=x_i$, $\tau_i=0$ and $\tau_{\!f}=\hbar\beta$, where $\beta=1/(k_BT)$, $k_B$ is Boltzmann's constant, and $T$ is the absolute temperature. The integral over $x(\tau)$ yields the trace of $e^{-\beta\hat{H}}$, which is the canonical partition function:
\begin{equation}
Z=\int\langle x|e^{-\beta\hat{H}}|x\rangle\,dx=\mbox{Tr}\bigl(e^{-\beta\hat{H}}\bigr)
=\int[Dx(\tau)]\exp\biggl[-{1\over\hbar}\oint_{0}^{\hbar\beta}L_E(x(\tau))\,d\tau\biggr]\, .
\label{eq10}
\end{equation}
The completeness relation,
\begin{equation}
\sum_{n=1}^\infty |n\rangle\langle n|=1\, ,
\label{eq11a}
\end{equation}
for the orthonormal eigenkets $\hat{H}|n\rangle=E_n|n\rangle$, when used in the trace in (\ref{eq10}), yields the partition function in the usual  form
\begin{equation}
Z=\sum_{n=0}^\infty e^{-\beta E_n}\, .
\end{equation}

\subsection{Correlation functions and propagators}
\label{sec2.3}

The energies of the ground state and excited states of the quantum anharmonic oscillator are encoded in correlation functions that are expectation values of products of the position operator:
\begin{equation}
\langle\hat{x}(\tau_1)\hat{x}(\tau_2)\cdots\hat{x}(\tau_n)\rangle={1\over Z}\mbox{Tr}\Bigl[e^{-\beta\hat{H}}\hat{x}(\tau_1)\hat{x}(\tau_2)\cdots\hat{x}(\tau_n)\Bigr]\, .
\label{eq12}
\end{equation}

The energy $E_0$ of the ground state can be obtained 
from the expectation of the Hamiltonian (\ref{eq1}) which, in conjunction with the virial theorem \cite{merzbacher98}, is expressed in terms of correlation functions of $x$ as \cite{creutz81}
\begin{equation}
E_0=m\omega^2\langle \hat{x}^2\rangle+3\lambda\langle \hat{x}^4\rangle\, .
\label{eq14a}
\end{equation}
An alternative expression for the group-state energy, more closely related to those for excited states derived below, is
\begin{equation}
-{d\log Z\over d\beta}=E_0\, .
\end{equation}

The correlation functions $\langle \hat{x}^n\rangle$ for $n=2$ and $4$ are, from (\ref{eq12}),
\begin{equation}
\langle x^n(\tau)\rangle={1\over Z}\mbox{Tr}\Bigl[e^{-\beta \hat{H}}\hat{x}^n(\tau)\Bigr]\, .
\end{equation}
By writing
\begin{equation}
\hat{x}(\tau)=e^{\hat{H}\tau/\hbar}\hat{x}(0)e^{-\hat{H}\tau/\hbar}\, ,
\label{eq15}
\end{equation}
which is the imaginary time counterpart of the Heisenberg picture for the time-dependence of operators,
again using (\ref{eq11a}), and taking the limit $\beta\to\infty$, we obtain
\begin{equation}
\langle x^n(\tau)\rangle=\langle 0|\hat{x}^n(0)|0\rangle\, .
\end{equation}
which is the expectation value of $\hat{x}^n$ in the ground state.

The two-point correlation function for calculating the first excited state is
\begin{equation}
\langle \hat{x}(\tau)\hat{x}(0)\rangle_c=\langle \hat{x}(\tau)\hat{x}(0)\rangle-\langle \hat{x}(\tau)\rangle\langle \hat{x}(0)\rangle\, ,
\label{eq17a}
\end{equation}
where the subscript ``$c$\,'' denotes cumulant, which in diagrammatic analysis correspond to a connected diagram. Again using (\ref{eq15}) and (\ref{eq11a}), yields
\begin{equation}
G_2(\tau)\equiv\lim_{\beta\to\infty}\langle \hat{x}(\tau)\hat{x}(0)\rangle_c
=\sum_{n=1}^\infty e^{-(E_n-E_0)\tau/\hbar}|\langle 0|\hat{x}|n\rangle|^2\, .
\label{eq17b}
\end{equation}
Hence, the energy of the first excited state is obtained as
\begin{equation}
-\hbar\lim_{\tau\to\infty}\biggl[{d\log G_2(\tau)\over d\tau}\biggr]=E_1-E_0\, .
\label{eq17c}
\end{equation}

The second excited state is obtained from the four-point connected correlation function
\begin{equation}
\langle\hat{x}(\tau)^2\hat{x}(0)^2\rangle_c=\langle\hat{x}(\tau)^2\hat{x}(0)^2\rangle-\langle\hat{x}(\tau)^2\rangle\langle\hat{x}(0)^2\rangle\, .
\end{equation}
By proceeding as above, we obtain
\begin{equation}
G_4(\tau)\equiv\lim_{\beta\to\infty}\langle\hat{x}(\tau)^2\hat{x}(0)^2\rangle_c
=\sum_{n=2}^\infty e^{-(E_n-E_0)\tau/\hbar}|\langle 0|\hat{x}|n\rangle|^2\, .
\end{equation}
The second excited state is, therefore, determined from
\begin{equation}
-\hbar\lim_{\beta\to\infty}\biggl[{d\log G_4(\tau)\over d\tau}\biggr]=E_2-E_0\, .
\label{eq22a}
\end{equation}

The probability density of the ground-state wave function is obtained by following the development of Creutz and Freedman \cite{creutz81}.  The probability $P(x)$ of a particle being found between positions $x-{1\over2}\Delta x$ and $x+{1\over2}\Delta x$ at any time $t^\prime$ in a real-time interval $[0,t]$ is given by the time-average
\begin{equation}
P(x)={1\over t}\int_0^t dt^\prime\int_{x-{1\over2}\Delta x}^{x+{1\over2}\Delta x}dx^\prime {\langle x_{\!f},t|x^\prime,t^\prime\rangle\langle x^\prime,t^\prime|x_i,0\rangle\over\langle x_{\!f},t|x_i,0\rangle}\, ,
\label{eq20a}
\end{equation}
The numerator in the integrand counts the paths that begin at  $(x_i,0)$, end at $(x_{\!f},t)$, and pass through $(x^\prime,t^\prime)$ for $0\le t^\prime\le t$. The propagator in the denominator counts all paths between $(x_i,0)$ and $(x_{\!f},t)$.

\begin{figure}[b]
\hskip-0.7cm\includegraphics[width=0.325\textwidth]{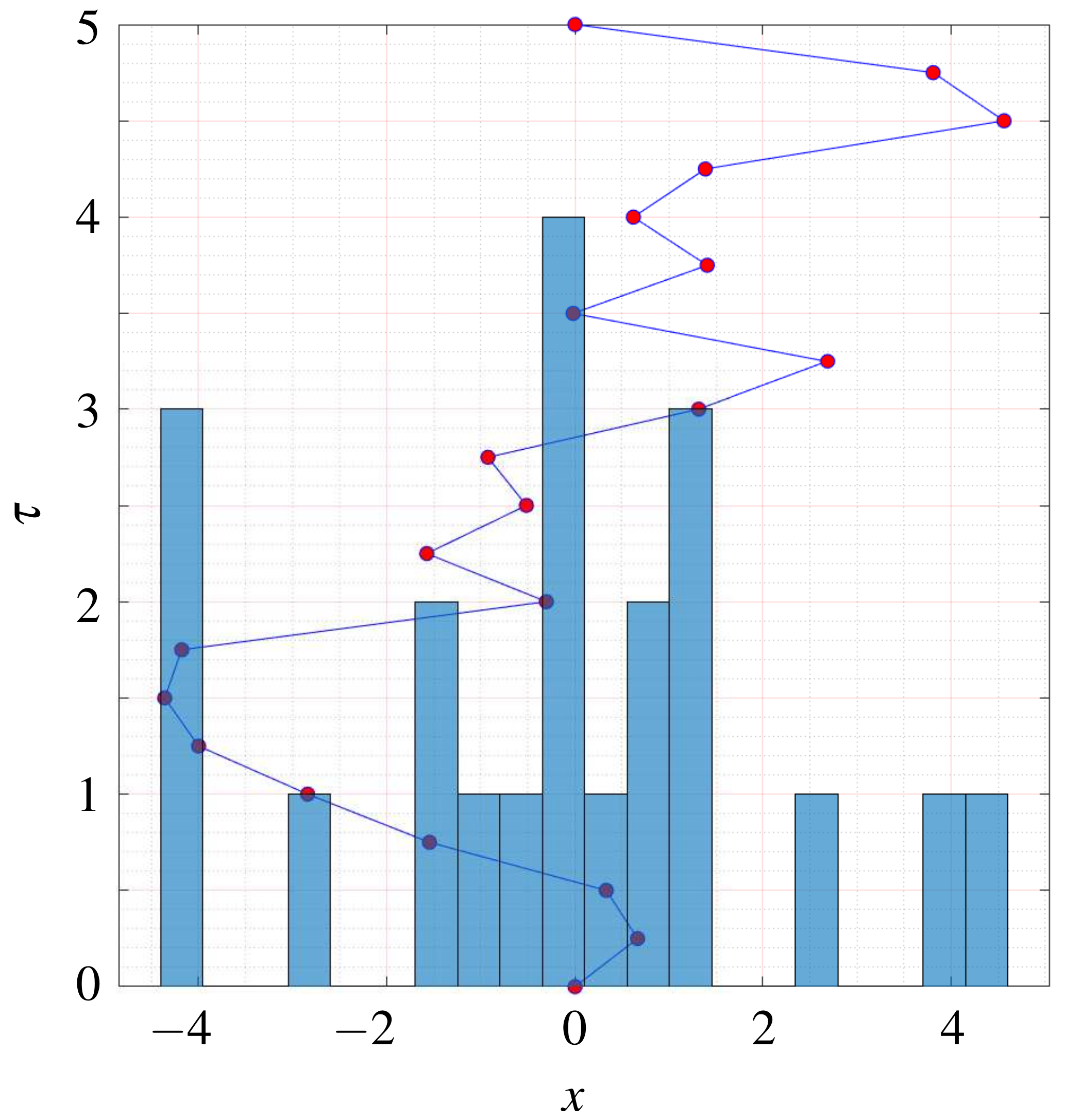}
\caption{A discrete imaginary-time path between $(0, 0)$ and $(0, 5)$. The histogram indicates the number of times the particle crosses the corresponding spatial region.}
\label{fig1}
\end{figure}

If $\Delta x$ is assumed to be small enough so that the integral over $x$ can be evaluated by keeping terms only to first order in $\Delta x$,  we obtain
\begin{equation}
P(x)={\Delta x\over t\langle x_{\!f},t|x_i,0\rangle}\int_0^t dt^\prime\langle x_{\!f},t|x,t^\prime\rangle\langle x,t^\prime|x_i,0\rangle\, .
\label{eq21a}
\end{equation}
The propagators are now written in terms of the eigenfunctions of $\hat{H}$ by using the completeness relation (\ref{eq11a})  for the eigenkets of $\hat{H}$:
\begin{equation}
\langle x^\prime, t^\prime|x,t\rangle=\langle x^\prime|e^{-i\hat{H}(t^\prime-t)/\hbar}|x\rangle
=\sum_{n=0}^\infty e^{-iE_n(t^\prime-t)/\hbar}\psi_n^\ast(x^\prime)\psi_n(x)\, .
\end{equation}
By using this expression for each propagator in (\ref{eq21a}) and continuing to imaginary time $\tau=\hbar\beta$, the long-imaginary-time/low-temperature limit yields the probability density of the ground state:
\begin{equation}
P(x)=|\psi_0(x)|^2\Delta x\, .
\label{eq23a}
\end{equation}
Figure~\ref{fig1} provides a schematic illustration of how the wave function is calculated by assigning a path along a time lattice to spatial bins with width $\Delta x$.

\section{Markov chain Monte Carlo method}
\label{sec3}

Monte Carlo simulations are carried out on a time lattice with $N_\tau$ time increments $\delta\tau$ with lattice points $x_n=n\delta\tau$ for $n=0,1,2,\ldots,N_\tau$. The $n$th time increment is $x_{n+1}-x_n$.  Periodic boundary conditions are imposed on the lattice, whereby increment $N_{\tau}+1$ equals increment $1$. The imaginary-time path integral in (\ref{eq8}), and (\ref{eq9}) is the continuum limit of the discretized expression
\begin{equation}
\langle x_{\!f}|e^{-\hat{H}(\tau_{\!f}-\tau_i)/\hbar}|x_i\rangle
\quad=\lim_{N_\tau\to\infty}\int\prod_{k=1}^{N_\tau} dx_k\biggl({m\over2\pi\hbar\,\delta\tau}\biggr)e^{-S(x_1,x_2,\ldots,x_n)/\hbar}\, ,
\label{eq20}
\end{equation}
where
\begin{equation}
S(\{x_k\})=\delta\tau\sum_{i=1}^{N_\tau}\biggl[{m\over2}\biggl({x_{i+1}-x_i\over\delta\tau}\biggr)^2+V(x_i)\biggr]\, .
\label{eq21}
\end{equation}
This is the Feynman-Kac formula \cite{feynman48,hibbs65,kac49}, which establishes a connection between the propagator (the Green's function for the Hamiltonian) and a path integral.

We work with a dimensionless action. With each variable expressed in terms of a suitable power of the lattice spacing $\delta\tau$, we work in units where $\hbar=1=c$, and introduce the dimensionless variables
\begin{equation}
\tilde{m}=m\delta\tau\, ,\qquad \tilde{\omega}=\omega\delta\tau\, ,\qquad \tilde{x}_i={x_i\over\delta\tau}\, .
\label{eq22}
\end{equation}
By combining (\ref{eq1}), (\ref{eq21}), and (\ref{eq22}), the action for the anharmonic oscillator becomes
\begin{equation}
S(\{x_k\})=\sum_{i=1}^{N_\tau}\biggl[{\tilde{m}\over2}(\tilde{x}_{i+1}-\tilde{x}_i)^2+{\tilde{m}\tilde{\omega}^2\tilde{x}_i^2\over2}+\tilde{\lambda}\tilde{m}^2\tilde{\omega}^3\tilde{x}_i^4\biggr]\, ,
\label{eq23}
\end{equation}
in which $\tilde{\lambda}$ is also dimensionless. 

The MCMC method is based on determining the statistics of observables from paths $(x_1,\ldots,x_{N_{\tau}})$ that are representative of the distribution in (\ref{eq20}) and (\ref{eq21}).  This requires generating reliable sequences of (pseudo) random numbers.  We have used the Mersenne twister \cite{mersenne98}.

We begin with an initial path $P^{(0)}$, which may be an array of random numbers (``hot'' start) or zeros (``cold'' start).  This path is updated by applying the Metropolis--Hastings algorithm \cite{metropolis53,hastings70} to each element $x_i$ of the path in random order, called a ``sweep.''  There are two steps in the updating process:

\begin{enumerate}

\item Generate a random number $u$ from a uniform distribution in the interval $[-h,h]$, where $h$, called the hit size, must be chosen judiciously. If $h$ is too large, few changes will be accepted; too small and the exploration of phase space will be slow.  For the sets of simulations here, hit sizes were chosen to obtain an acceptance rate of 50-60\%. 

\item Propose the new value, $x_i^\prime=x_i+u$, of the path element and calculate the resulting change $\Delta S$ in the action. New values that lower the action are always accepted, while those that would increase the action are accepted with probability $e^{-\Delta S}$.

\end{enumerate}

One sweep produces the next path, e.g.~$P^{(1)}$ from $P^{(0)}$.  Each path is determined only by the immediately preceding path, so the complete sequence of paths forms a Markov chain, but the paths are correlated. The accuracy of the MCMC method relies on sampling from the distribution in (\ref{eq20}) and (\ref{eq21}) which, in turn, relies on the paths being stationary and independent. The initial path ``thermalizes'', that is, attained equilibrium after $N_{\rm therm}$ sweeps. To counteract the inherent autocorrelation in a Markov chain, a number $N_{\rm sep}$ of paths between successive paths used for measurements (i.e.~representative of the equilibrium distribution) must be discarded.  A detailed description of this process is provided in Ref.~\onlinecite{westbroek18b}.

\section{Parameters for MCMC simulations}
\label{sec4}

\begin{figure}[t!]
\hskip-0.7cm\includegraphics[width=0.325\textwidth]{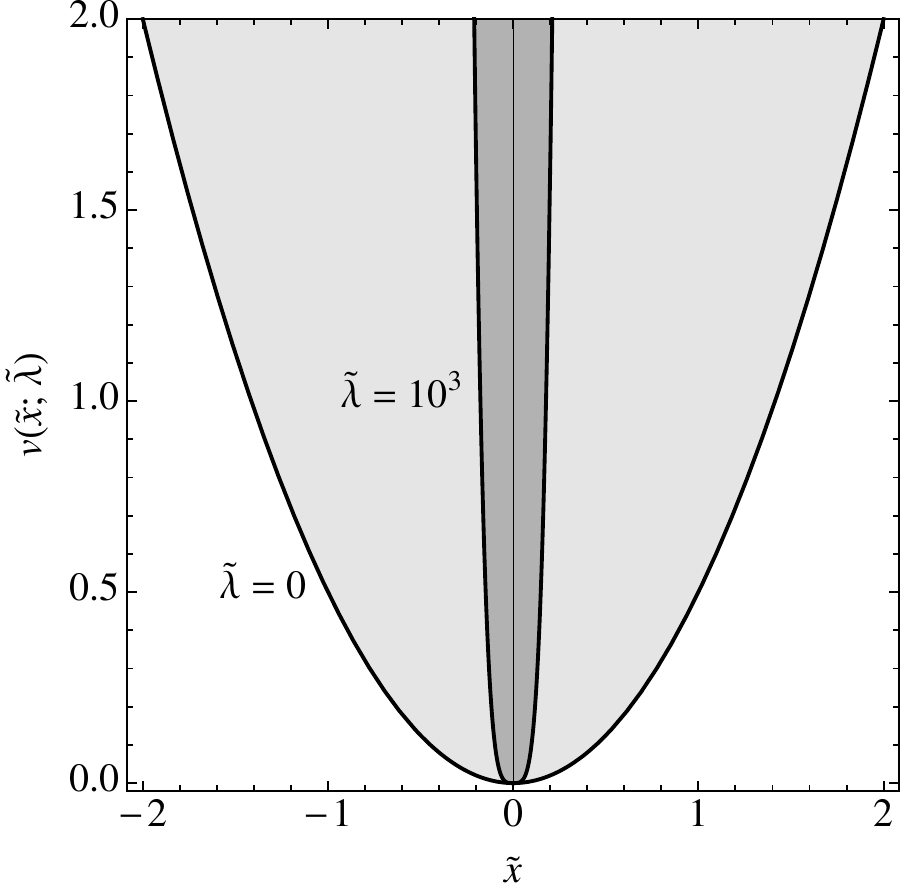}
\caption{The potential $v(x;\lambda)={1\over2}\tilde{x}^2+\tilde{\lambda}\tilde{x}^4$ for $\tilde{\lambda}=0$, which is the potential for the  harmonic oscillator (light shading), and $\tilde{\lambda}=10^3$, which corresponds to the strong quartic limit (dark shading). The quartic potential localizes the wave functions of the oscillators, which causes the corresponding energy eigenvalues to increase.}
\label{fig2}
\end{figure}

For all calculations reported here, $\tilde{m}=\tilde{\omega}=\delta\tau$ and $N_\tau\delta\tau=250$ in (\ref{eq23}). Calculations have been performed  for $\tilde{\lambda}=0,1,50$ and $10^3$, which range from the harmonic oscillator to the strong quartic limit (Fig.~\ref{fig2}).  Such a large variation of $\tilde{\lambda}$ affects not just the quantum mechanical behavior of the oscillator, but also several parameters used in the MCMC method:~the hit size $h$ needed to achieve an acceptance rate of 50-60\%, and the number $N_{\rm sep}$ of paths that must be discarded between successive paths used for calculations. The coupling constant also affects the convergence to the continuum limit of the probability density of the ground state and the energy levels.  

\begin{figure}[t!]
\centering
\includegraphics[width=0.475\textwidth]{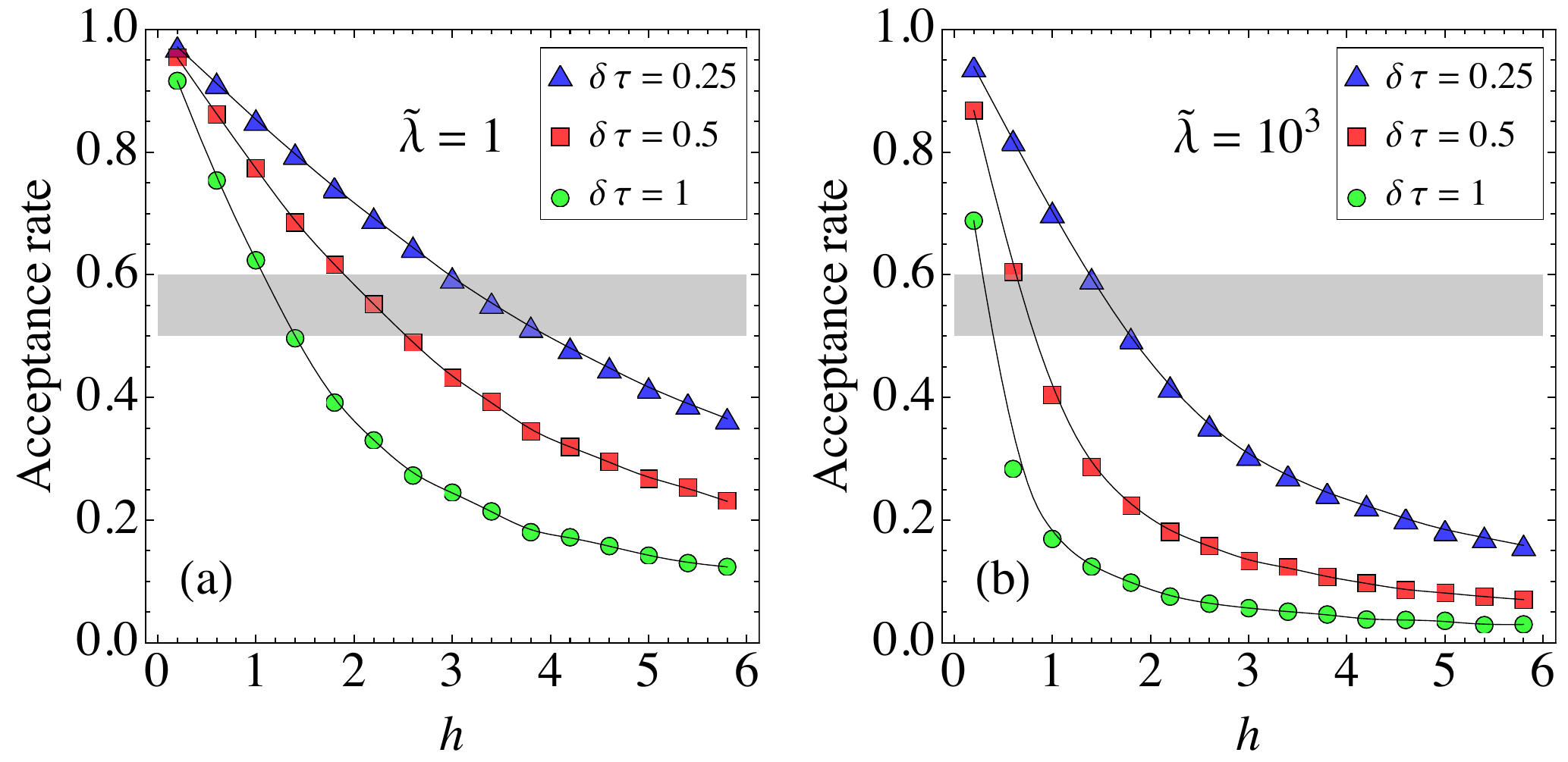}
\caption{Acceptance rate versus the hit size $h$ for the quantum anharmonic oscillator with quartic coupling constant (a) $\tilde{\lambda}=1$ and (b) $\tilde{\lambda}=1000$ for the indicated discretizations.  The target of 50-60\% acceptance rate is indicated by shading.  The curves are spline fits to the data.}
\label{fig3}
\end{figure}

Figure~\ref{fig3} compares the acceptance rate versus hit size for the anharmonic oscillator with a weak $(\tilde{\lambda}=1)$ and a strong $(\tilde{\lambda}=10^3)$ quartic coupling constant for several discretizations.  These data were obtained as follows:~(i) begin with a cold start and $h=0.2$, (ii) ignore several initial sweeps, (iii) calculate the acceptance rate, as given in the pseudocode in Ref.~\onlinecite{westbroek18b}, (iv) calculate the arithmetic mean of the acceptance rate for every 100 sweeps.  The hit size is then increased by 0.2 and the process (i)--(iv) is repeated.

Most apparent in Fig.~\ref{fig3} is that, for each discretization, the target hit size is much smaller for the larger quartic coupling constant, which  leads to larger changes in the action with increasing coupling and, thus, suppresses the acceptance rate. In other words, the exploration of phase space is slower, which is a natural consequence of the more localized potential associated with a stronger coupling constant (Fig. \ref{fig2}). 

A key element of the MCMC method is the selection of paths used for calculations.  These paths must be representative of the equilibrium distribution, as defined by the partition function, so the paths must first equilibrate from some initial configuration. The number $N_{\rm therm}$ of sweeps required to attain equilibrium is determined when the measured quantity fluctuates about a steady state, which depends on the observable, but generally increases as either $\delta\tau$ or $\tilde{\lambda}$ decreases. Typically, the equilibration of several observables is plotted and the maximum number of sweeps to equilibrium is used for all simulations.

\begin{figure}[t]
\centering
\includegraphics[width=0.475\textwidth]{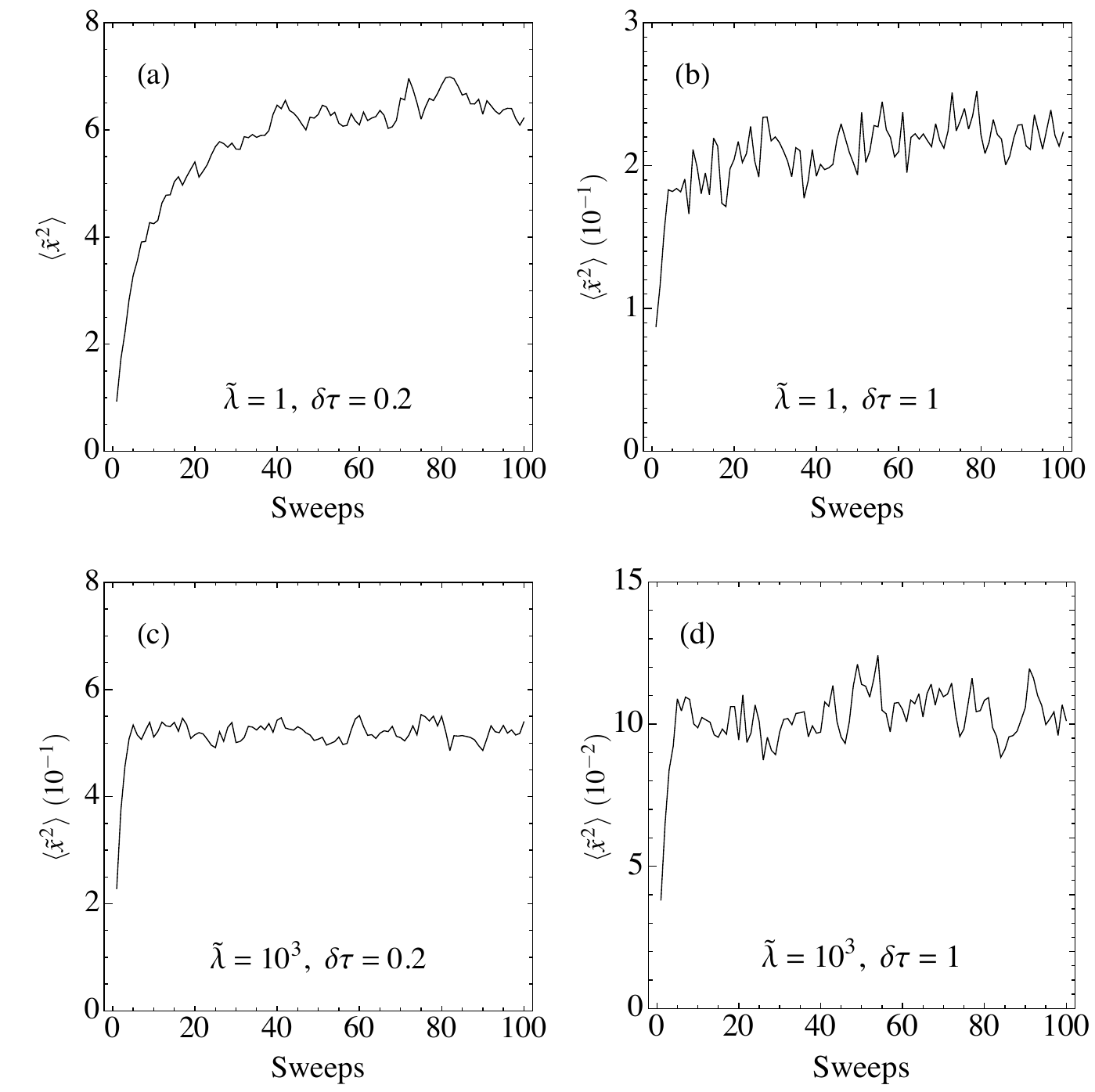}
\caption{Equilibration of $\langle\tilde{x}^2\rangle$ over the first 100 sweeps from the initial configuration for (a) $\tilde{\lambda}=1$ and $\delta\tau=0.2$,  (b) $\tilde{\lambda}=1$ and $\delta\tau=1$, (c) $\tilde{\lambda}=10^3$ and $\delta\tau=0.2$, and (d) $\tilde{\lambda}=10^3$ and $\delta\tau=1$, with $N_\tau\delta\tau=250$ for all simulations.  Note the differences in the scales along the vertical axis.}
\label{fig4}
\end{figure}

Figure~\ref{fig4} shows the equilibration for $\langle\tilde{x}^2\rangle$ for the anharmonic oscillator with coupling constants $\tilde{\lambda}=1$ (a,b) and $\tilde{\lambda}=10^3$ (c,d) for lattice spacings $\delta\tau=0.2$ (a,c) and $\delta\tau=1$ (b,d).  Note the differences in scales of $\langle\tilde{x}^2\rangle$ in each panel, which shows that the equilibrium value is much smaller for the system with the larger coupling constant.  Also immediately apparent is how many fewer sweeps from the initial path are required to attain equilibrium with increasing coupling constant.

The trends in Figs.~\ref{fig3} and \ref{fig4} can be explained by the sharpening of the potential energy profile with increasing $\tilde{\lambda}$ (Fig.~\ref{fig2}).  All of our simulations began with cold starts.  Thus, with increasing $\lambda$, the accessible configuration space decreases, so the attainment of equilibrium is correspondingly quicker. Similarly, the acceptance rate for a fixed hit size $h$ is lower for larger $\tilde{\lambda}$ because of the increasing confinement near the origin, at the expense of the classically forbidden region. This also explains why, for an observable such as $\langle\tilde{x}^2\rangle$ (Fig.~\ref{fig4}), sweeps over Markov chains show smaller fluctuations around their mean for larger $\tilde{\lambda}$.  Finally, the increasing equilibrium value of $\langle\tilde{x}^2\rangle$ with $\tilde{\lambda}$ is a direct result of the increasingly localized probability density of the ground-state wave function (Sec.~\ref{sec5.1}) caused by the sharpening potential energy profile (Fig.~\ref{fig2}).  

The number $N_{\rm sep}$ of sweeps discarded between successive measurements, and the hit size $h$ for the discretizations $\delta\tau$ used for the calculations described in the following section are compiled in Table~\ref{table1}. These entries were obtained from the data in Figs.~\ref{fig3} and \ref{fig4} for each discretization and quartic coupling constant.  The variations of simulation parameters in this table extends to other calculations using the MCMC method, such as the autocorrelation times and the application of the jackknife analysis for analyzing the statistics of correlated samples.  All of our simulations began with cold starts, and we used 200 paths (every 100th sample out of a chain of 20,000) for all measurements. We used the method and code of Ref.~\onlinecite{wolff04} for estimating autocorrelation times and error analysis.

\begin{table}[t]
\caption{\label{table1} 
The initial $N_{\rm therm}$ Metropolis--Hastings sweeps that are neglected and the hit size $h$, presented as $(N_{\rm therm},h)$, for the indicated discretizations $\delta\tau$ and coupling constants $\tilde{\lambda}$.  For these simulations, $N_{\rm therm}$ and $N_{\rm sep}$ were of similar order of magnitude. No simulations were performed for values of $\tilde{\lambda}$ and $\delta\tau$ whose entry is indicated by a dash (--).}
\begin{ruledtabular}
\begin{tabular}{cllll}
$\delta\tau$&\multicolumn{1}{c}{$\tilde{\lambda}=0$}&\multicolumn{1}{c}{$\tilde{\lambda}=1$}&\multicolumn{1}{c}{$\tilde{\lambda}=50$}& \multicolumn{1}{c}{$\tilde{\lambda}=1000$}\\\hline
0.01&\multicolumn{1}{c}{--}&\multicolumn{1}{c}{--}&\multicolumn{1}{c}{--}&200,\,16\\
0.02&\multicolumn{1}{c}{--}&\multicolumn{1}{c}{--}&300,\,11&100,\,14\\
0.05&\multicolumn{1}{c}{--}&500,\,9&100,\,9&100,\,8\\
0.10&500,\,5&200,\,5&100,\,5&100,\,6\\
0.20&100,\,4&100,\,4&100,\,3&100,\,2\\
0.25&100.\,3.5&100,\,3.5&100,\,2.5&100,\,1.5\\
0.40&100,\,2.5&100,\,3.5&100,\,1.6&100,\,0.9\\
0.50&100,\,2.5&100,\,2&100,\,1.3&100,\,0.7\\
1.00&100,\,1.5&100,\,1.2&100,\,0.5&100,\,0.3
\end{tabular}
\end{ruledtabular}
\end{table}

\section{The Quantum Anharmonic Oscillator}
\label{sec5}

\subsection{Probability density of the ground state}
\label{sec5.1}

The properties of our Markov chains enable an indirect evaluation of the long-imaginary-time limit (\ref{eq23}) of the expression (\ref{eq20}) for the probability density of the ground state.  In particular, the Markov chains in Sec.~\ref{sec3}, which are aperiodic, irreducible, and positively recurrent, guarantee not only that any initial chain approaches equilibrium, as the discussion accompanying Fig.~\ref{fig4} demonstrates, but that long-time limits may be replaced by ensemble averages (ergodicity) \cite{morningstar07,Roberts}.

\begin{figure}[t!]
\centering
\includegraphics[width=0.475\textwidth]{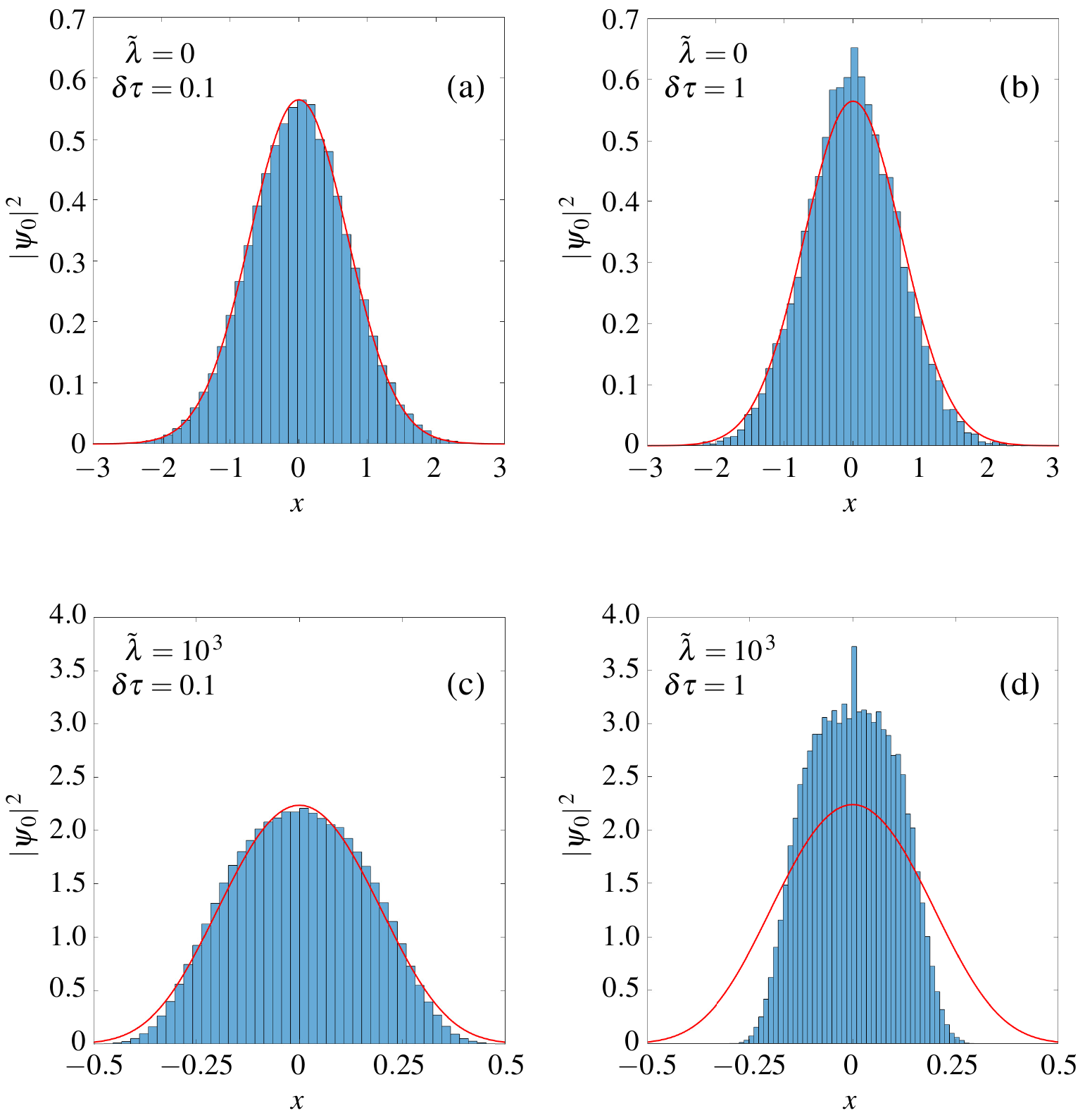}
\caption{The probability densities $|\psi_0(x)|^2$ of the ground-state wave functions for the harmonic oscillator (a,b) and the strong quartic limit (c,d) in the Schr\"odinger equation with Hamiltonian (\ref{eq23}) for $\delta\tau=1$ (b,d) and $\delta\tau=0.1$ (a,c).  The histograms were obtained according to the procedure shown in Fig.~\ref{fig1}, with 200 paths (every 100th from a chain of 20000) used for determining the probability density. The red curves superimposed on these histograms are numerical solutions to Schr\"odinger's equations with the corresponding Hamiltonians (\ref{eq23}) obtained by using the {\tt bvp4c} solver of MATLAB (Ref.~\onlinecite{matlab,kutz13}).}
\label{fig5}
\end{figure}

Figure~\ref{fig5} shows the (normalized) probability densities $|\psi_0(x)|^2$ of the wave functions for the ground state of the Hamiltonian (\ref{eq23}) with $\tilde{\lambda}=0$, which is the harmonic oscillator, and with $\tilde{\lambda}=10^3$, in which the quartic term dominates.  The numerical integration of the corresponding Schr\"odinger equations is shown for comparison.  Perhaps the most striking aspect of this figure is the enhanced localization of the wave function for $\tilde{\lambda}=10^3$ compared with that for the harmonic oscillator ($\tilde{\lambda}=0$).  This is to be expected from the narrowing of the potential with increasing $\tilde{\lambda}$ (Fig.~\ref{fig1}). Moreover, the rate of convergence to the continuum limit is considerably slower for the large quartic term.  The errors for $\delta\tau=1$ are small for the harmonic oscillator, but are substantial for the large quartic potential, with small discrepancies remaining in the tail of the distribution even for $\delta\tau=0.1$.

\subsection{Ground-state energy}
\label{sec5.2}

\begin{figure}[b!]
\centering
\includegraphics[width=0.48\textwidth]{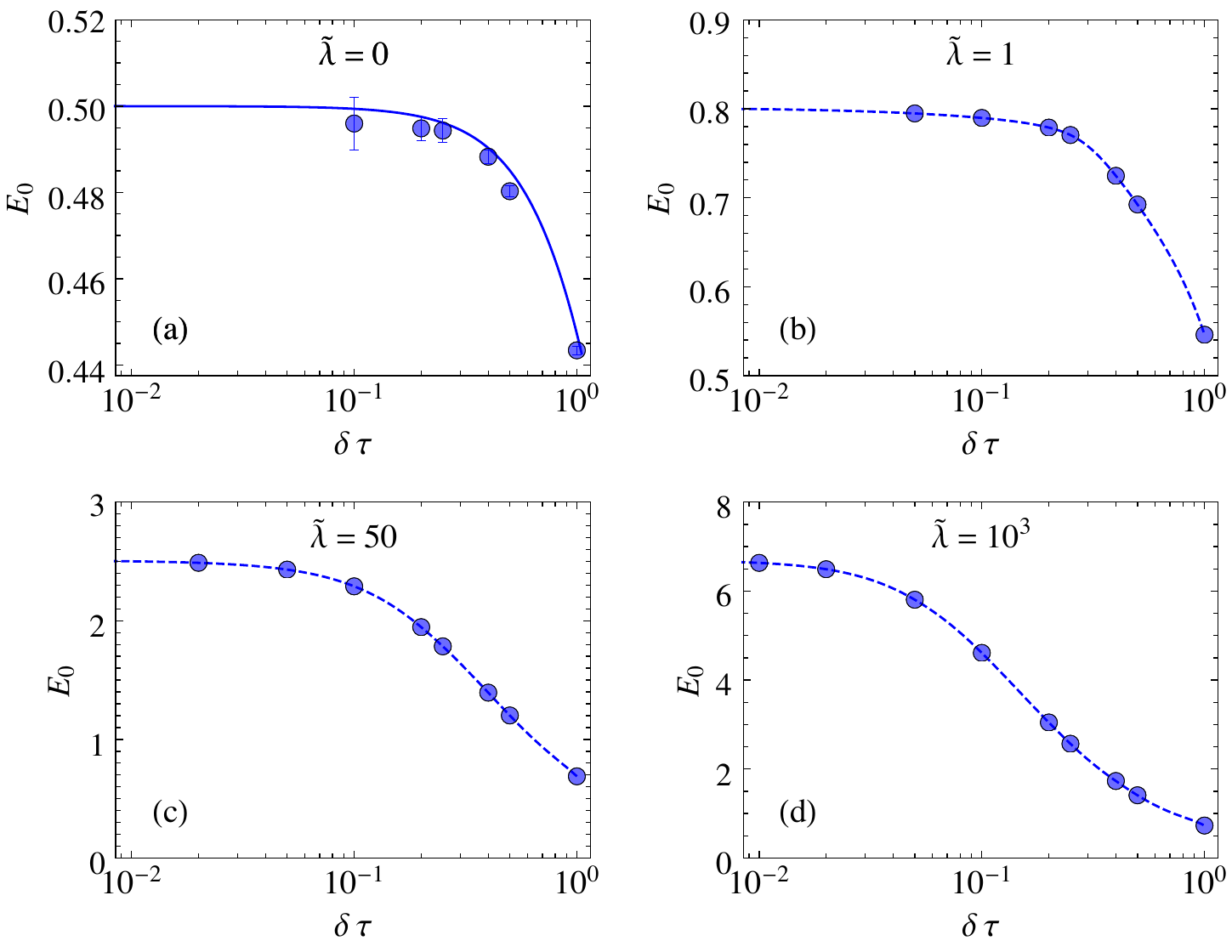}
\caption{Calculation of the ground-state energy of the anharmonic oscillator with quartic couplings (a) $\tilde{\lambda}=0$, (b) $\tilde{\lambda}=1$, (c) $\tilde{\lambda}=50$, and (d) $\tilde{\lambda}=10^3$.  The filled circles represent values calculated from the MCMC method. In (a), the solid line is the exact result calculated in Ref.~\onlinecite{creutz81,westbroek18a}, while in (b-d), the broken curve is a (not-a-knot) cubic spline fit carried out on linear axes.  The logarithmic axis for $\delta\tau$ is for presentation purposes only. Where error bars are not indicated, the errors are of the same size or smaller than the symbol.}
\label{fig6}
\end{figure}

The energies of anharmonic oscillators were expressed as correlation functions in Sec.~\ref{sec2.3}.   The MCMC method necessitates evaluating discrete approximations to these quantities, which are then used to estimate the energies at decreasing lattice spacing $\delta\tau$.  The results with the simulation parameters in Table~\ref{table1} are shown in Fig.~\ref{fig6} for $\tilde{\lambda}=0$, $1$, $50$, and $10^3$.  Also shown are the exact results \cite{creutz81,westbroek18a} for the harmonic oscillator (Fig.~\ref{fig6}(a)) and cubic spline fits to the data points for the anharmonic systems (Fig.~\ref{fig6}(b,c,d)). The spline fits were used to estimate the ground-state energies in the continuum limit on linear axes;~the logarithmic axis for $\delta\tau$ is used in Fig.~\ref{fig6} for presentation purposes only.

The ground-state energies for the systems shown in Fig.~\ref{fig6} have been calculated by several methods, with the results compiled in Table~\ref{table2}.  The energies in the column labelled MCMC are obtained from the finest discretization in Fig.~\ref{fig6}, while the column Spline labels the energies extrapolated from the spline fits to the calculated data points.  These two columns are the results obtained from the path integral method, either directly (MCMC) or inferred (Spline).  The remaining two columns contain essentially exact numerical results, obtained by the numerical integration of Schr\"odinger's equation, and from the method of Hioe and Montroll \cite{hioe75}, who used the Bargmann representation to develop rapidly converging algorithms for the energy levels of oscillators as a function of the anharmonic coupling constant.

\begin{table}[t!]
\caption{\label{table2}Ground-state energy of the anharmonic oscillator for the indicated values of $\tilde{\lambda}$.  The column labelled MCMC is the data point in Fig.~\ref{fig6} with the finest discretization, Spline is the value obtained by extrapolating the spline curve, SE is the energy obtained by the numerical integration of Schr\"odinger's equation, and the last column contains the energies calculated by the method in Ref.~\onlinecite{hioe75} for $\tilde{\lambda}>0$.}
\begin{ruledtabular}
\begin{tabular}{cdddd}
$\tilde{\lambda}$&\multicolumn{1}{c}{MCMC}&\multicolumn{1}{c}{Spline}&\multicolumn{1}{c}{SE}& \multicolumn{1}{c}{Ref.~\onlinecite{hioe75}}\\\hline
0&\multicolumn{1}{c}{0.496}&\multicolumn{1}{c}{0.501}&\multicolumn{1}{c}{${1\over2}$}&\multicolumn{1}{c}{--}\\
1&\multicolumn{1}{c}{0.795}&\multicolumn{1}{c}{0.801}&\multicolumn{1}{c}{0.8038}&\multicolumn{1}{c}{0.8038}\\
50&\multicolumn{1}{c}{2.488}&\multicolumn{1}{c}{2.511}&\multicolumn{1}{c}{2.4998}&\multicolumn{1}{c}{2.4997}\\
$10^3$&\multicolumn{1}{c}{6.634}&\multicolumn{1}{c}{6.702}&\multicolumn{1}{c}{6.6941}&\multicolumn{1}{c}{6.6942}
\end{tabular}
\end{ruledtabular}
\end{table}

The ground-state energy obtained from the extrapolation of the spline are within a few tenths of a per cent of the exact results.  In contrast, the energies obtained from the path integral with the finest discretization show discrepancies by as much as 9\%.  This highlights the dual role of the spline fit:~(i) as an estimate of the energy by extrapolation, and (ii) an indication of the improvement expected by reducing the time lattice spacing.

\subsection{First excited state}
\label{sec5.3}

\begin{figure}[t!]
\hskip-0.7cm\includegraphics[width=0.35\textwidth]{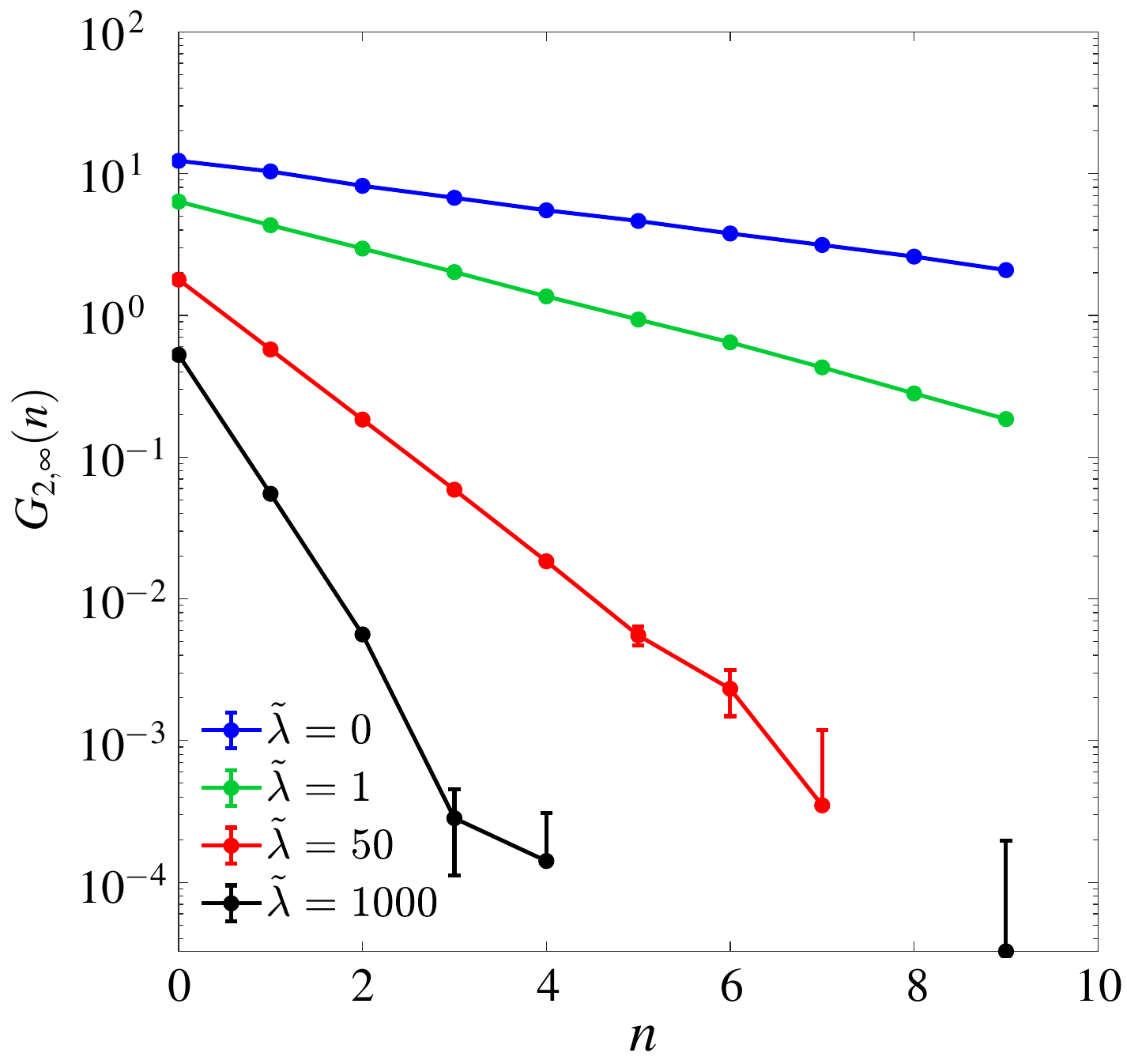}
\caption{The asymptotic form of two-point correlation function in (\ref{eq32}) for the indicated values of $\tilde{\lambda}$ for a lattice spacing of $\delta\tau=0.2$. The slope for small values of $n$ determine the difference $E_1-E_0$ between the ground state and the first excited state.}
\label{fig7}
\end{figure}

The first excited of the anharmonic oscillator is obtained by evaluating the logarithmic derivative of the long-time limit of $G(\tau)$ in (\ref{eq17b}), which is the zero-temperature limit of the correlation function in (\ref{eq17a}). Equation (\ref{eq17c}) then establishes the difference $E_1-E_0$ between the ground state and first excited state as the negative of the slope of $G(\tau)$ in the long-$\tau$ limit. On a time lattice, (\ref{eq17c}) is approximated as
\begin{equation}
-{(E_1-E_0)\over\hbar}\approx\lim_{\tau\to\infty}\biggl\{{\log[G_2(\tau+\Delta\tau)]-\log[G_2(\tau)]\over\Delta\tau}\biggr\}
=\lim_{\tau\to\infty}\biggl\{{1\over\Delta\tau}\log\biggl[{G_2(\tau+\Delta\tau)\over G_2(\tau)}\biggr]\biggr\}\, .
\end{equation}
The approximate solution of this equation,
\begin{equation}
G_{2,\infty}(\Delta\tau)=G_{2,\infty}(0)e^{-(E_1-E_0)\Delta\tau/\hbar}\, ,
\label{eq32}
\end{equation}
is independent of $\tau$. As $\Delta\tau=n\delta\tau$ for some non-negative integer $n$, we can determine $E_1-E_0$ by plotting $\log[G_{2,\infty}(n)]$ versus $n$.  An example is shown in Fig.~\ref{fig7} for $\tilde{\lambda}=0$, $1$, $50$, and $10^3$ with $\delta\tau=0.2$. The linear behavior, evident for small $n$, enables estimates to be made for $E_1$, given the values of $E_0$ in Table~\ref{table2}.  

\begin{figure}[t!]
\centering
\includegraphics[width=0.48\textwidth]{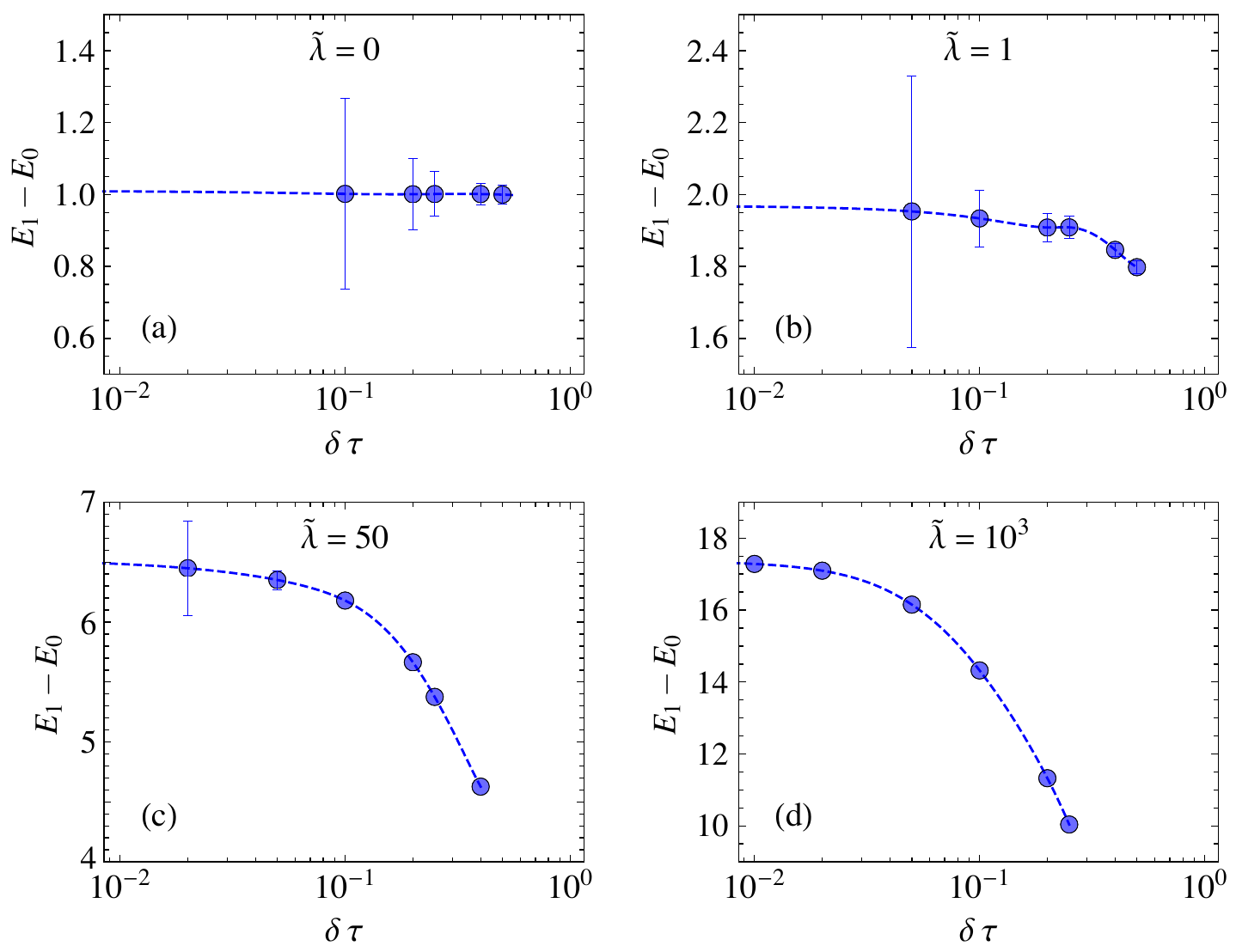}
\caption{Calculation of the energy difference $E_1-E_0$ of the anharmonic oscillator with quartic couplings (a) $\tilde{\lambda}=0$, (b) $\tilde{\lambda}=1$, (c) $\tilde{\lambda}=50$, and (d) $\tilde{\lambda}=10^3$.  The filled circles represent values calculated from the MCMC method. The broken curves are a (not-a-knot) cubic spline fits carried out on linear axes.  The logarithmic axis for $\delta\tau$ is for presentation purposes only. Where error bars are not indicated, the errors are smaller than the symbol size.}
\label{fig8}
\end{figure}

The calculation of the energy differences $E_1-E_0$ are shown in Fig.~\ref{fig8} and the resulting energies of the first excited states are shown in Table~\ref{table3}. The differences between estimates based on the extrapolation of the cubic spline and the exact values are of the order of 1\% or less.  However, the error bars for these calculations, particularly for $\tilde{\lambda}=0$ and $\tilde{\lambda}=1$, are much larger than the corresponding calculations of the ground-state energies (Fig.~\ref{fig6}). Note, in particular, that despite these large error bars, the mean of each calculation for the harmonic oscillator $(\tilde{\lambda}=0)$ is close to the exact value of 1.

\begin{table}[b!]
\caption{\label{table3}Energy of the first excited state $E_1$ of anharmonic oscillators for the indicated values of $\tilde{\lambda}$.  The first shows the energy differences $E_1-E_0$ obtained from the extrapolation of the spline fits in Fig.~\ref{fig8}. The column Spline is $E_1$ obtained from extrapolating spline curves in Figs.~\ref{fig6} and \ref{fig8}, SE is $E_1$ obtained by the numerical integration of Schr\"odinger's equation, and the last column contains the energies calculated by the method in Ref.~\onlinecite{hioe75} for $\tilde{\lambda}>0$.}
\begin{ruledtabular}
\begin{tabular}{cdddd}
$\tilde{\lambda}$&\multicolumn{1}{c}{$E_1-E_0$}&\multicolumn{1}{c}{Spline $(E_1)$}&\multicolumn{1}{c}{SE $(E_1)$}& \multicolumn{1}{c}{Ref.~\onlinecite{hioe75} $(E_1)$}\\\hline
0&\multicolumn{1}{c}{1.010}&\multicolumn{1}{c}{1.511}&\multicolumn{1}{c}{${3\over2}$}&\multicolumn{1}{c}{--}\\
1&\multicolumn{1}{c}{1.969}&\multicolumn{1}{c}{2.770}&\multicolumn{1}{c}{2.7379}&\multicolumn{1}{c}{2.7379}\\
50&\multicolumn{1}{c}{6.523}&\multicolumn{1}{c}{9.034}&\multicolumn{1}{c}{8.9155}&\multicolumn{1}{c}{8.9151}\\
$10^3$&\multicolumn{1}{c}{17.368}&\multicolumn{1}{c}{24.069}&\multicolumn{1}{c}{23.9731}&\multicolumn{1}{c}{23.9722}
\end{tabular}
\end{ruledtabular}
\end{table}

\subsection{Second excited state}
\label{sec5.4}

\begin{figure}[t!]
\centering
\includegraphics[width=0.48\textwidth]{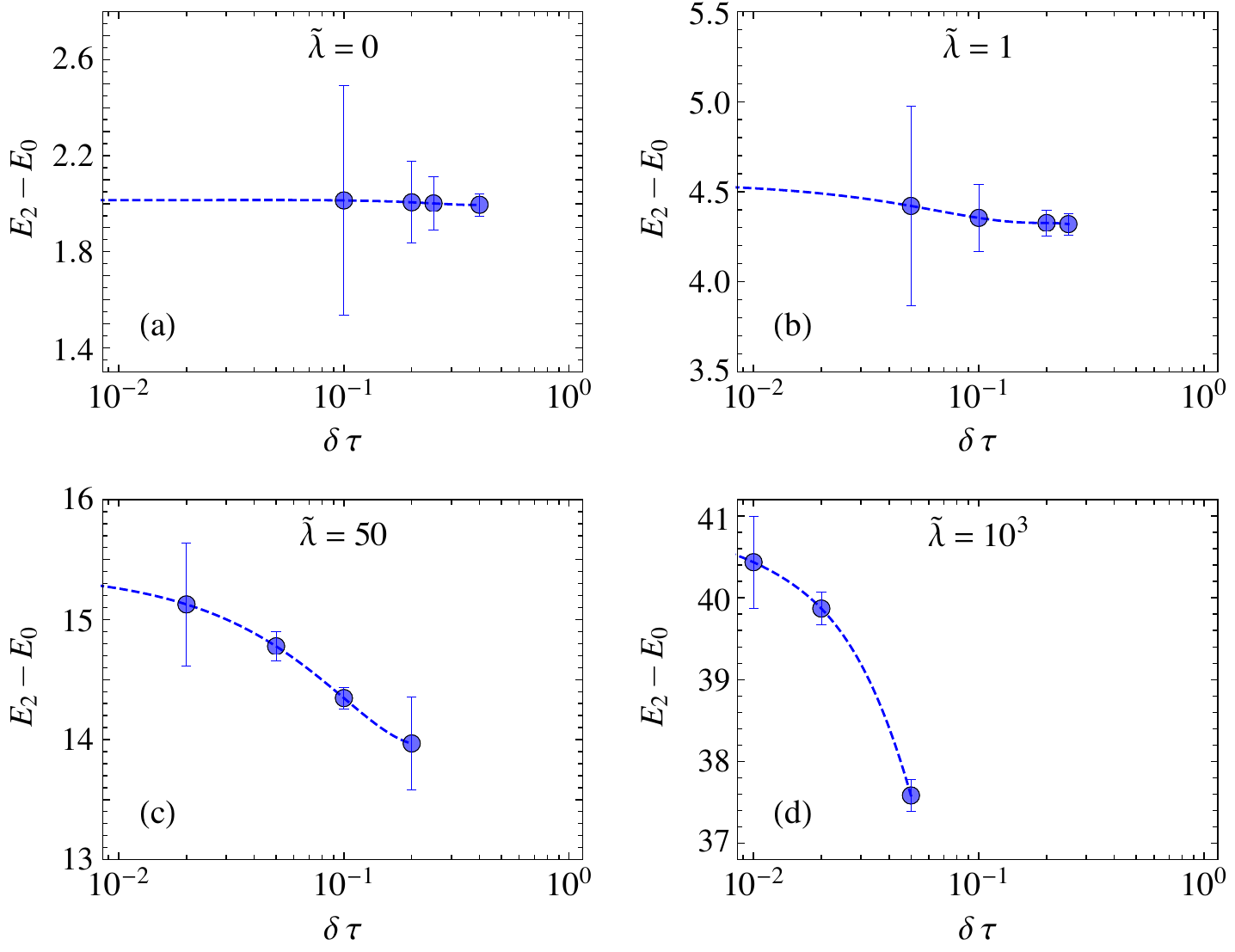}
\caption{Calculation of the energy difference $E_2-E_0$ of the anharmonic oscillator with quartic couplings (a) $\tilde{\lambda}=0$, (b) $\tilde{\lambda}=1$, (c) $\tilde{\lambda}=50$, and (d) $\tilde{\lambda}=10^3$.  The filled circles represent values calculated from the MCMC method. The broken curves are a (not-a-knot) cubic spline fits carried out on linear axes.  The logarithmic axis for $\delta\tau$ is for presentation purposes only. Where error bars are not indicated, the errors are smaller than the symbol size.}
\label{fig9}
\end{figure}

The determination of the second excited state from (\ref{eq22a}) proceeds in a manner similar to that in the preceding section.  The approximate solution analogous to (\ref{eq32}) is
\begin{equation}
G_{4,\infty}(\Delta\tau)=G_{4,\infty}(0)e^{-(E_2-E_0)\Delta\tau/\hbar}\, ,
\end{equation}
where $G_{4,\infty}$ is the approximate solution of 
\begin{equation}
-{(E_2-E_0)\over\hbar}\approx\lim_{\tau\to\infty}\biggl\{{1\over\Delta\tau}\log\biggl[{G_2(\tau+\Delta\tau)\over G_2(\tau)}\biggr]\biggr\}\, .
\end{equation}
and is again independent of $\tau$.  The energy difference $E_2-E_0$ is obtained by plotting $\log[G_{4,\infty}(n)]$ versus $n$.

The results of such an analysis are shown in Fig.~\ref{fig9}. The larger error bars for all of the oscillators are clearly evident, as is the more limited range of discretizations that are practical.  Nevertheless, when estimates for $E_2-E_0$ obtained from the extrapolation of the spline fits in Fig.~\ref{fig9} are combined with the corresponding values of $E_0$ in Fig.~\ref{fig6} are compared with exact calculations, our estimates are found to agree to within a few per cent (Table~\ref{table4}).

\begin{table}[b!]
\caption{\label{table4}Energy of the second excited state $E_2$ of anharmonic oscillators for the indicated values of $\tilde{\lambda}$.  The first shows the energy differences $E_2-E_0$ obtained from the extrapolation of the spline fits in Fig.~\ref{fig9}. The column Spline is $E_2$ obtained from extrapolating spline curves in Figs.~\ref{fig6} and \ref{fig9}, SE is $E_2$ obtained by the numerical integration of Schr\"odinger's equation, and the last column contains the energies calculated by the method in Ref.~\onlinecite{hioe75} for $\tilde{\lambda}>0$.}
\begin{ruledtabular}
\begin{tabular}{cdddd}
$\tilde{\lambda}$&\multicolumn{1}{c}{$E_2-E_0$}&\multicolumn{1}{c}{Spline $(E_2)$}&\multicolumn{1}{c}{SE $(E_2)$}& \multicolumn{1}{c}{Ref.~\onlinecite{hioe75} $(E_2)$}\\\hline
0&\multicolumn{1}{c}{2.014}&\multicolumn{1}{c}{2.515}&\multicolumn{1}{c}{${5\over2}$}&\multicolumn{1}{c}{--}\\
1&\multicolumn{1}{c}{4.551}&\multicolumn{1}{c}{5.352}&\multicolumn{1}{c}{5.1794}&\multicolumn{1}{c}{5.1793}\\
50&\multicolumn{1}{c}{15.400}&\multicolumn{1}{c}{17.911}&\multicolumn{1}{c}{17.4379}&\multicolumn{1}{c}{17.4370}\\
$10^3$&\multicolumn{1}{c}{40.904}&\multicolumn{1}{c}{47.606}&\multicolumn{1}{c}{47.0202}&\multicolumn{1}{c}{47.0173}
\end{tabular}
\end{ruledtabular}
\end{table}

\section{Summary and Outlook}

We have applied the MCMC method to the evaluation of the imaginary-time path integral for the anharmonic oscillator with a coupling strengths ranging from the harmonic limit to strongly quartic.  Quantities calculated include the probability density of the ground state and the energies of the ground state and the first two excited states. The ground-state probability density becomes more localized as the potential becomes more localized due to the increase in the quartic coupling constant, and the energy levels increase accordingly, as expected from elementary quantum mechanics. Information about higher lying excited states becomes increasingly difficult to extract because they are exponentially suppressed in the imaginary-time path integral.  Indeed, a comparison of Figs.~\ref{fig6}, \ref{fig8}, and \ref{fig9} illustrates how the statistical fluctuations become more pronounced away from the ground state for the same number of paths used for measurements.  We will present calculations of higher excited states in a separate publication.

Nevertheless, the imaginary-time path integral formulation has many attractive features for applications in condensed matter and other fields in physics.  The method we have used here, which is described in detail in Ref.~\onlinecite{creutz81,westbroek18b}, can be applied to a particle in any potential.  The inclusion of interactions between several particles is also within the scope of the method and is of interest even in one-dimensional systems.  For example, the behavior of electrons in quantum wires and carbon nanotubes, where electronic motion is allowed in one dimension, but strongly restricted in the two lateral dimensions, is a central concern in theoretical and experimental studies of these systems.  The harmonic potential can be used as a model of electrons (and holes) in semiconductor quantum dots.

\begin{acknowledgments}
MJEW was supported through a Janet Watson scholarship from the Department of Earth Science and Engineering and a studentship in the Centre for Doctoral Training on Theory and Simulation of Materials funded by the EPSRC (EP/L015579/1).
\end{acknowledgments}

\appendix

\section{The Partition Function}
\label{sec1}
\noindent
The partition function is
\begin{equation}
Z=\mbox{Tr}\left(e^{-\beta\hat{H}}\right)\, .
\end{equation}
Inserting the completeness relation for the orthonormal eigenkets of $\hat{H}$, $1=\sum_{n=0}^\infty|n\rangle\langle n|$, into the partition function yields
\begin{align}
Z&=\mbox{Tr}\left[e^{-\beta\hat{H}}\left(\sum_{n=0}^\infty\ket{n}\bra{n}\right)\right]=\mbox{Tr}\left(\sum_{n=0}^\infty\ket{n} e^{-\beta E_n}\bra{n}\right)\nonumber\\
\noalign{\vskip3pt}
&=\sum_{n=0}^\infty e^{-\beta E_n}\bra{n}\ket{n}=\sum_{n=0}^\infty e^{-\beta E_n}\, ,
\end{align}
which is the more familiar form of the partition function.
\section{Correlation Functions}\label{sec2}
\noindent
Consider the $2n$-point correlation function:
\begin{equation}
G_{2n}(\tau)=\langle \hat{x}^{\,n}(\tau)\hat{x}^{\,n}(0)\rangle-\langle\hat{x}^{\,n}(\tau)\rangle\langle\hat{x}^{\,n}(0)\rangle\, .
\label{eq2.1}
\end{equation}
The first term is defined as
\begin{align}
\langle \hat{x}^{\,n}(\tau)\hat{x}^{\,n}(0)\rangle&={1\over Z}\mbox{Tr}\left[e^{-\beta\hat{H}}\hat{x}^{\,n}(\tau)\hat{x}^{\,n}(0)\right]\nonumber\\
\noalign{\vskip3pt}
&={1\over Z}\mbox{Tr}\left[e^{-\beta\hat{H}}e^{\hat{H}\tau/\hbar}\hat{x}^{\,n}(0)e^{-\hat{H}\tau/\hbar}\hat{x}^{\,n}(0)\right]\, .
\end{align}
Using the completeness relation for the orthonormal eigenkets of $\hat{H}$ twice in the right-hand side,
\begin{align}
\langle \hat{x}^{\,n}(\tau)\hat{x}^{\,n}(0)\rangle&={1\over Z}\mbox{Tr}\left[e^{-\beta\hat{H}}e^{\hat{H}\tau/\hbar}\left(\sum_{p=0}^\infty\ket{p}\bra{p}\right)\hat{x}^{\,n}(0)e^{-\hat{H}\tau/\hbar}\left(\sum_{q=0}^\infty\ket{q}\bra{q}\right)\hat{x}^{\,n}(0)\right]\nonumber\\
\noalign{\vskip3pt}
&={1\over Z}\mbox{Tr}\left[\sum_{p=0}^\infty\sum_{q=0}^\infty\ket{p} e^{-\beta E_p}e^{E_p\tau/\hbar}\bra{p}\hat{x}^{\,n}(0)\ket{q} e^{-E_q\tau/\hbar}\bra{q}\hat{x}^{\,n}(0)\right]\nonumber\\
\noalign{\vskip3pt}
&={1\over Z}\sum_{p=0}^\infty\sum_{q=0}^\infty e^{-\beta E_p}e^{E_p\tau/\hbar}\bra{p}\hat{x}^{\,n}(0)\ket{q} e^{-E_q\tau/\hbar}\bra{q}\hat{x}^{\,n}(0)\ket{p}\nonumber\\
\noalign{\vskip3pt}
&={1\over Z}\sum_{p=0}^\infty\sum_{q=0}^\infty e^{-\beta E_p}e^{-(E_q-E_p)\tau/\hbar}\bigl|\bra{p}\hat{x}^n(0)\ket{q}\bigr|^2\, .\label{eq2.3}
\end{align}
We also have
\begin{equation}
\langle \hat{x}^{\,n}(\tau)\rangle={1\over Z}\mbox{Tr}\left[e^{-\beta\hat{H}}\hat{x}^{\,n}(\tau)\right]={1\over Z}\mbox{Tr}\left[e^{-\beta\hat{H}}e^{\hat{H}\tau/\hbar}\hat{x}^{\,n}(0)e^{-\hat{H}\tau/\hbar}\right]\, .
\end{equation}
Again using the completeness relation twice in the right-hand side,
\begin{align}
\langle \hat{x}^{\,n}(\tau)\rangle&={1\over Z}\mbox{Tr}\left[e^{-\beta\hat{H}}e^{\hat{H}\tau/\hbar}\left(\sum_{p=0}^\infty\ket{p}\bra{p}\right)\hat{x}^{\,n}(0)e^{-\hat{H}\tau/\hbar}\left(\sum_{q=0}^\infty\ket{q}\bra{q}\right)\right]\nonumber\\
\noalign{\vskip3pt}
&={1\over Z}\mbox{Tr}\left[\sum_{p=0}^\infty\sum_{q=0}^\infty\ket{p} e^{-\beta E_p}e^{E_p\tau/\hbar}\bra{p}\hat{x}^{\,n}(0)\ket{q} e^{-E_q\tau/\hbar}\bra{q}\right]\nonumber\\
\noalign{\vskip3pt}
&={1\over Z}\sum_{p=0}^\infty\sum_{q=0}^\infty e^{-\beta E_p}e^{E_p\tau/\hbar}\bra{p}\hat{x}^{\,n}(0)\ket{q} e^{-E_q\tau/\hbar}\bra{q}\ket{p}\nonumber\\
\noalign{\vskip3pt}
&={1\over Z}\sum_{p=0}^\infty e^{-\beta E_p}\bra{p}\hat{x}^{\,n}(0)\ket{p}\, .
\label{eq2.5}
\end{align}
Since the expression for $\langle \hat{x}^{\,n}(\tau)\rangle$ does not depend on $\tau$ (which is expected) 
\begin{equation}
\langle \hat{x}^n(0)\rangle={1\over Z}\sum_{p=0}^\infty e^{-\beta E_p}\bra{p}\hat{x}^{\,n}(0)\ket{p}\, ,\label{eq2.6}
\end{equation}
which is the same as \eqref{eq2.5}. Thus, by combining \eqref{eq2.1}, \eqref{eq2.3}, \eqref{eq2.5} and \eqref{eq2.6}, we obtain
\begin{align}
G_{2n}(\tau)&={1\over Z}\sum_{p=0}^\infty\sum_{q=0}^\infty e^{-\beta E_p}e^{-(E_q-E_p)\tau/\hbar}\bigl|\bra{p}\hat{x}^n(0)\ket{q}\bigr|^2\ldots\nonumber\\
\noalign{\vskip3pt}
&\quad-\left[{1\over Z}\sum_{p=0}^\infty e^{-\beta E_p}\bra{p}\hat{x}^{\,n}(0)\ket{p}\right]\left[{1\over Z}\sum_{p=0}^\infty e^{-\beta E_p}\bra{p}\hat{x}^n(0)\ket{p}\right]\, .\label{eq2.7}
\end{align}
\noindent
To take the low-temperature limit of $G_{2n}$, we first have that\cite{reif85}
\begin{equation}
Z=\sum_{n=0}^\infty e^{-\beta E_n}\, .\label{eq10}
\end{equation}
Then, for the terms in \eqref{eq2.7}, the limit $\beta\to\infty$ yields
\begin{align}
\lim_{\beta\to\infty}\left({1\over Z}\sum_{p=0}^\infty e^{-\beta E_p}\bra{p}\hat{x}^n(0)\ket{p}\right)&=\lim_{\beta\to\infty}\left({e^{-\beta E_0}\bra{0}\hat{x}^n(0)\ket{0}+e^{-\beta E_1}\bra{1}\hat{x}^n(0)\ket{1}+\cdots\over e^{-\beta E_0}+e^{-\beta E_1}+\cdots}\right)\nonumber\\
\noalign{\vskip3pt}
&=\lim_{\beta\to\infty}\left({\bra{0}\hat{x}^n(0)\ket{0}+e^{-\beta(E_1-E_0)}\bra{1}\hat{x}^n(0)\ket{1}+\cdots\over 1+e^{-\beta(E_1-E_0)}+\cdots}\right)\nonumber\\
\noalign{\vskip3pt}
&=\bra{0}\hat{x}^n(0)\ket{0}\, .
\end{align}
Similarly,
\begin{align}
&\lim_{\beta\to\infty}\left({1\over Z}\sum_{p=0}^\infty\sum_{q=0}^\infty e^{-\beta E_p}e^{-(E_q-E_p)\tau/\hbar}\bigl|\bra{p}\hat{x}^n(0)\ket{q}\bigr|^2\right)\nonumber\\
\noalign{\vskip3pt}
&=\lim_{\beta\to\infty}\left({\sum_{q=0}^\infty e^{-\beta E_0} e^{-(E_q-E_0)\tau/\hbar}\bigl|\bra{0}\hat{x}^n(0)\ket{q}\bigr|^2+
e^{-\beta E_1} e^{-(E_q-E_1)\tau/\hbar}\bigl|\bra{1}\hat{x}^n(0)\ket{q}\bigr|^2+\cdots\over e^{-\beta E_0}+e^{-\beta E_1}+\cdots}\right)\nonumber\\
\noalign{\vskip3pt}
&=\lim_{\beta\to\infty}\left({\sum_{q=0}^\infty  e^{-(E_q-E_0)\tau/\hbar}\bigl|\bra{0}\hat{x}^n(0)\ket{q}\bigr|^2+
e^{-\beta (E_1-E_0)} e^{-(E_q-E_1)\tau/\hbar}\bigl|\bra{1}\hat{x}^n(0)\ket{q}\bigr|^2+\cdots\over 1+e^{-\beta (E_1-E_0)}+\cdots}\right)\nonumber\\
\noalign{\vskip3pt}
&=\sum_{q=0}^\infty  e^{-(E_q-E_0)\tau/\hbar}\bigl|\bra{0}\hat{x}^n(0)\ket{q}\bigr|^2\, .
\end{align}
Hence,
\begin{equation}
G_{2n}(\tau)=\sum_{q=0}^\infty  e^{-(E_q-E_0)\tau/\hbar}\bigl|\bra{0}\hat{x}^n(0)\ket{q}\bigr|^2-\bigl|\bra{0}\hat{x}^n(0)\ket{0}\bigr|^2\, .\label{eq11}
\end{equation}
For $n=1$, 
\begin{align}
G_{2}(\tau)&=\sum_{q=0}^\infty  e^{-(E_q-E_0)\tau/\hbar}\bigl|\bra{0}\hat{x}(0)\ket{q}\bigr|^2-\bigl|\bra{0}\hat{x}(0)\ket{0}\bigr|^2\nonumber\\
&=\sum_{q=1}^\infty  e^{-(E_q-E_0)\tau/\hbar}\bigl|\bra{0}\hat{x}(0)\ket{q}\bigr|^2\, .
\end{align}
For $n=2$,
\begin{align}
G_{4}(\tau)&=\sum_{q=0}^\infty  e^{-(E_q-E_0)\tau/\hbar}\bigl|\bra{0}\hat{x}^2(0)\ket{q}\bigr|^2-\bigl|\bra{0}\hat{x}^2(0)\ket{0}\bigr|^2\nonumber\\
&=\sum_{q=2}^\infty  e^{-(E_q-E_0)\tau/\hbar}\bigl|\bra{0}\hat{x}^2(0)\ket{q}\bigr|^2\, .
\label{eq2.13}
\end{align}
Consider the matrix element $\langle 0|\hat{x}^2(0)|q\rangle$. For the harmonic oscillator, the ground state has even parity, the first excited state has odd parity, and parity alternates between even and odd for all higher lying states.  We expect the same pattern for the anharmonic oscillator.  Therefore, as the operator $\hat{x}^2$ and the ground-state wave function are both even functions of $x$, the matrix element $\langle 0|\hat{x}^2(0)|q\rangle$ vanishes if the wave function of the $q$th excited state is an odd function of $x$.  In particular $\bra{0}\hat{x}^2(0)\ket{1}=0$.  Hence, the sum in \eqref{eq2.13} begins with $q=2$.

\section{Monte Carlo Markov Chains}
\label{sec3}

\subsection{Background}

The sum over paths in the evaluation of the path integral is carried out with the Markov chain Monte Carlo (MCMC) method.  This method was developed by Metropolis {\it  et al.}, \cite{metropolis53} who were studying the equilibrium properties of interacting particle systems.  Instead of simulating the actual dynamical relaxation of such a system to equilibrium, the authors made the key observation that, for the calculation of equilibrium properties, a Markov chain that attains the same equilibrium distribution by any sequence of intermediate states is sufficient.  Hasting\cite{hastings70} viewed the Metropolis method mainly as a way of sampling high-dimensional probability distributions, which is, in fact, the modern use of this algorithm. Hasting's article was written in a statistical style that abstracted the Metropolis method into a transition operator on a Markov chain whose target distribution is the invariant distribution of the chain, rather than the equilibrium distribution of Metropolis method.  Simulations that follow this approach are said to use the {\it Metropolis--Hastings} algorithm.

\subsection{Basic Concepts and Notation}
\label{sec3.0}

A Markov chain is a sequence of events in which a given position in the chain depends only on the immediately preceding position, rather than on the history of the chain.  A Markov chain is defined by three attributes:

\begin{enumerate}

\item A {\it state space} which defines the values  that can be taken by the chain.

\item A {\it transition operator}  that defines the probability of one state progressing to another state.

\item An {\it initial condition} that specifies the state from which the progression of the chain begins.

\end{enumerate}

Our goal is to calculate the expectation value $\langle\hat{\cal O}\rangle$ of some operator ${\cal O}$ over a finite time interval,
\begin{equation}
\langle\hat{\cal O}\rangle={\sum_{k=1}^N{\cal O}(\bm{x}_k)}P_{\rm eq}(\bm{x}_k)\, ,
\label{eq16a}
\end{equation}
where $\bm{x}_k=\bigl(x_1^{(k)},x_2^{(k)},\ldots,x_N^{(k)}\bigr)$ is a configuration of the system, which also represents a path, and $P_{\rm eq}(\bm{x}_k)$ is the probability distribution in the canonical ensemble. The MCMC solution to this problem to the calculation of (\ref{eq16a}) is to construct a Markov chain on the state space $\bm{X}$ that has $P_{\rm eq}(\bm{x})$ as the stationary distribution.  In other words, given transition probabilities $W(\bm{x},\bm{x}^\prime)$ for chains $\bm{x}$ and $\bm{x}^\prime$ in state space, we have
\begin{equation}
\int P_{\rm eq}(\bm{x})W(\bm{x},\bm{x}^\prime)\,d\bm{x}=P_{\rm eq}(\bm{x}^\prime)\, .
\label{eq18a}
\end{equation}

The discretized path integral representation of the canonical partition function in \eqref{eq10}, 
\begin{align}
Z&=\int\prod_{k=1}^{N_\tau} dx_k\biggl({m\over2\pi\hbar\,\delta\tau}\biggr)\exp\left\{-{\delta\tau\over\hbar}
\sum_{i=1}^{N_\tau}\left[{m\over2}\biggl({x_{i+1}-x_i\over\delta\tau}\biggr)^2+V(x_i)\right]\right\}\nonumber\\
\noalign{\vskip3pt}
&\equiv\int [D\bm{x}(\tau)]e^{-S(\bm{x})/\hbar}\, ,
\end{align}
identifies the equilibrium probability of the configuration $\bm{x}=\{x_1,x_2,\ldots,x_{N_\tau}\}$ as
\begin{equation}
P_{\rm eq}(\bm{x})={e^{-S(\bm{x})\hbar}\over Z}\, ,
\end{equation}
and averages of an operator ${\cal O}(\bm{x})$ are calculated as
\begin{equation}
\langle{\cal O}\rangle=\int_X[D\bm{x}(\tau)]{\cal O}(\bm{x})P_{\rm eq}(\bm{x})\, ,
\label{eq26a}
\end{equation}
in which $[D\bm{x}(\tau)]$ signifies that the integral is over all paths in the state space $X$.

The Monte Carlo method introduced by Metropolis is based on the idea of ``importance sampling'', whereby the phase-space points $\bm{x}$ in \eqref{eq26a} are not selected completely at random, which would be impractical for higher-dimensional spaces, but are chosen to be more densely distributed in region(s) of phase space providing the dominant contributions to the integral. The ergodicity of the Markov chains used here enables us to calculate averages as arithmetic averages over Markov chains:
\begin{equation}
\int_X[D\bm{x}(\tau)]{\cal O}(\bm{x})P_{\rm eq}(\bm{x})=\lim_{n\to\infty}\left[{1\over n}\sum_{i=1}^n{\cal O}(\bm{x}_i)\right]\, ,
\label{eq27a}
\end{equation}
where $n$ is the number of states generated by the Markov chain.  Of course, in actual calculations, only a finite number of states are used to estimate averages, as determined by the operator and the system.  The next section provides the mathematical foundation for \eqref{eq27a}.

\subsection{Properties of Markov Chains}
\label{sec3.1}

If the state space is finite or countable, then the elements of the Markov chain can be represented as vectors and the transition probability as a matrix that operates on these vectors.  However, most applications of the MCMC method have uncountable state spaces in which the initial state is an unconditional probability distribution and the transitions are expressed in terms of a conditional probability distribution.  Finite state spaces are simpler to present so, we will use this setting in what follows.

For a Markov chain that consists of $n$ states, the transition operator is an $n\times n$ matrix ${\sf P}$ whose entries $p_{ij}$ signify the probability of moving from state $s_i$ to state $s_j$ in a single step. Similarly, the probability to move from state $s_i$ to $s_j$ in 2 steps is $\sum_{k=1}^np_{ik}p_{kj}$, which is the $(i,j)$th element of ${\sf P}^2$. The generalization to $m$ steps is the $(i,j)$th element of ${\sf P}^m$.

If a transition operator does not change across transitions, the Markov chain is called (time) homogeneous.  An important consequence of homogeneity is that, as $t \rightarrow \infty$, the Markov chain will reach an equilibrium called the stationary distribution of the chain. The stationary distribution of a Markov chain is important for sampling from probability distributions, which is at the heart of MCMC methods. In more formal terms, suppose that the current state of the Markov chain is represented by the probability vector $\boldsymbol{u}$ which means that $u_i$ is the probability of being in state $s_i$. Now we want to know the probability $v_j$ of finding the chain in state $s_j$ after $m$-steps. In matrix form, this is given by 
\begin{equation}
\boldsymbol{v}^{\tt T}=\boldsymbol{u}^{\tt T}{\sf P}^m\, .
\end{equation}
If there is a vector such that $\boldsymbol{w}^{\tt T}=\boldsymbol{w}^{\tt T}{\sf P}$, which also implies $\boldsymbol{w}^{\tt T}=\boldsymbol{w}^{\tt T}{\sf P}^m$, then the Markov chain is said to be in \textit{equilibrium} and $\boldsymbol{w}$ is called a \textit{stationary} vector. Computationally, to find such a vector, one calculates the left eigenstate of the matrix ${\sf P}$ corresponding to unit eigenvalue.

The continuous state analogue of $p_{ij}$ is the transition density $p(x,y)$.  The continuous quantity corresponding to the $m$-step transition probability ${\sf P}^m$ is denoted as $p^{(m)}(x,y)$, which is defined by the Chapman--Kolmogorov recursion relation,
\begin{equation}
p^{(n)}(x,y)=\int_S p^{(n-1)}(x,z)p(z,y)\,dz\, ,
\end{equation}
for $n\ge2$.

The most important theorem behind the MCMC method is:\cite{morningstar07, Roberts}
\medskip

\noindent
\textbf{Fundamental limit theorem}. An irreducible, aperiodic Markov chain with transition matrix ${\sf P}$ has a stationary distribution $\boldsymbol{w}$ satisfying $w_j>0$, $\sum_jw_j=1$, and $\boldsymbol{w}^{\tt T}=\boldsymbol{w}^{\tt T}{\sf P}$, if and only if all its states are positively recurrent, and $w$ is unique and identical to the limiting distribution
\begin{equation}
w_j=\lim_{n\rightarrow\infty}({\sf P}^n)_{ij}
\label{eq16}
\end{equation}
with the following definitions:

\begin{itemize}

\item An \textit{irreducible} Markov chain is the one in which the probability to go from any state $s_i$ to any other state $s_j$ in a finite number of steps is non-zero. Thus, there exists a finite $n$ such that
\begin{equation}
\sum_{k_1}\sum_{k_2}\ldots\sum_{k_{n-1}}p_{ik_1}p_{k_1k_2}\ldots p_{k_{n-1}j}\neq 0\, .
\end{equation}

\item The states of a Markov chain are called \emph{positive recurrent} if the probability to return to the same state is unity after a finite number of evolution steps.

\item The states are \emph{periodic} if the transition $s_i\rightarrow s_i$ is only possible in steps which are integral multiples of the period $d(i)$. The period is defined as the highest common factor for all $m$ for which $({\sf P}^m)_{ii}>0$. For an aperiodic state $s_j$, $d(j)=1$.
\end{itemize}

Note that the stationary distribution obtained in \eqref{eq16} is independent of initial state $s_i$. This theorem gives us the freedom to start in any \emph{ergodic} (aperiodic, irreducible and positive recurrent) Markov chain and we are guaranteed to end up in an equilibrium state. The process of evolution of the chain into a stationary state is called \emph{thermalization}.  Although stated in terms of discrete states, with suitable modifications, this theorem is also valid for continuous states.  We refer the reader to Ref.~\onlinecite{Roberts} for details.

\section{Splines}
\label{sec4}

There are several methods for fitting curves to a set of given data points that enable the prediction of values between these points (interpolation) and the estimate of values outside the range of the points (extrapolation).  In spline interpolation, piecewise functions pass through a set of data points, often referred to as {\it knots}, such that the function is smooth at these data points and satisfies some specified conditions at or near the first and last points. Thus, if there are $n$ points $(x_i,y_i)$, for $i=1,2,\ldots,n$, a total of $n-1$ functions is used, one for each interval between the $i$th and $(i+1)$st data points, for $i=1,2,\ldots,n-1$. Cubic polynomials are most commonly employed to strike a balance between smoothness and avoiding Runge's phenomenon, which is the appearance of oscillations at the edges of an interval that occurs when using high-order polynomials for a set of equi-distant interpolation points. The cubic polynomials are written as
\begin{equation}
P_i(x)=a_i(x-x_i)^3+b_i(x-x_i)^2+c_i(x-x_i)+d_i\, ,
\end{equation}
where $i=1,2,\ldots,n-1$ for $x\in[x_i,x_{i+1}]$.  Each polynomial has four unknown coefficients $(a_i,b_i,c_i,d_i)$, so the complete spline has $4(n-1)$ unknowns. These unknowns must be chosen according to conditions that the spline must satisfy:

\begin{enumerate}

\item $P_i(x_i)=y_i$ and $P_{n-1}(x_n)=y_n$ for $i=1,2,\ldots,n-1$.  These conditions guarantee that the spline interpolates the data points.

\item $P_i(x_i)=P_{i+1}(x_i)$ for $i=1,2,\ldots,n-1$. These conditions require that the values of adjacent polynomials are the same at the points where they meet, which ensure that the interpolating spline is {\it continuous}.

\item $\displaystyle{{dP_{i-1}\over dx}\bigg|_{x=x_i}={dP_i\over dx}\bigg|_{x=x_i}}$  for $i=2,\ldots,n-1$ or, in abbreviated notation, $P_{i-1}^\prime(x_i)=P_i^\prime(x_i)$. These conditions supplement the continuity of the spline by requiring that the slopes of adjacent polynomials are the same at the point where they meet.  Thus, the spline is {\it differentiable} on $(x_1,x_n)$.  

\item $\displaystyle{{d^2P_{i-1}\over dx^2}\bigg|_{x=x_i}={d^2P_i\over dx^2}\bigg|_{x=x_i}}$  for $i=2,\ldots,n-1$ or, in abbreviated notation, $P_{i-1}^{\prime\prime}(x_i)=P_i^{\prime\prime}(x_i)$.  This condition supplements the continuity and differentiability of the spline by requiring that the second derivatives of adjacent polynomials are the same at the point where they meet.  Thus, the spline has a {\it second derivative} at every point on $(x_1,x_n)$.  

\end{enumerate}

\noindent
The four conditions on the spline provide 
\begin{equation}
(n-1)+(n-1)+(n-2)+(n-2)=4n-6
\end{equation}
constraints on the $4(n-1)$ coefficients of the spline. That leaves $4n-4-(4n-6)=2$ conditions left to uniquely determine the spline. These are imposed at or near the two end points of the data set, effectively providing boundary conditions for the spline function.  There are various ways of specifying these boundary conditions:

\begin{enumerate}

\item The {\it natural spline} boundary conditions are
\begin{equation}
{d^2P_1\over dx^2}\bigg|_{x=x_1}={d^2P_{n-1}\over dx^2}\bigg|_{x=x_{n}}=0,
\end{equation}
that is, the second derivatives vanishes at the end points.  These boundary conditions seldom used since they does not provide a sufficiently
accurate approximation near the end points of the data set.

\item The {\it clamped spline} boundary conditions set the first derivatives at the end points to a particular value, which may be known, or specified at the user's discretion:
\begin{equation}
{dP_1\over dx}\bigg|_{x=x_1}=f_1^\prime\, ,\qquad {dP_{n-1}\over dx}\bigg|_{x=x_{n}}=f_n^\prime\, .
\end{equation}
The equations determining the coefficients for cubic splines for either natural or clamped boundary conditions can be expressed as tridiagonal matrices.\cite{press}

\item The {\it not-a-knot} boundary conditions, where no extra conditions are imposed at the end points. At each point (i.e.~knot), the spline changes from the cubic polynomial in the preceding interval changes smoothly to the polynomial in the next interval. The basic idea of the not-a-knot boundary conditions is to not change the cubic polynomials across the second and penultimate points, which are first two interior points. Thus, these two points are not knots, which results in the first two intervals having the same spline function and the last two intervals having the same two spline functions.  The mathematical expression of these boundary conditions are
\begin{equation}
{d^3P_1\over dx^3}\bigg|_{x=x_2}={d^3P_2\over dx^3}\bigg|_{x=x_2}\, ,\qquad 
{d^3P_{n-2}\over dx^3}\bigg|_{x=x_{n-1}}={d^3P_{n-1}\over dx^3}\bigg|_{x=x_{n-1}}\, .
\end{equation}
The default in-built MATLAB function performs a cubic not-a-knot spline fit.

\end{enumerate}

\section{Computer Programs}
\label{sec5}

Two MATLAB codes were used to produce the results in our research. The first one is numerical evaluation of the path integral using MCMC. Within this code we have used another short MATLAB code \verb|UWerr.m|, which is freely available on the internet.\cite{wolff04} The purpose of \verb|UWerr.m| is to calculate the mean of a sample which may exhibit autocorrelations. The second code is the direct integration of the Schr\"odinger equation. The following are the codes, created on MATLAB R2018a Update 3 (9.4.0.885841).
\begin{lstlisting}[style= Matlab-editor,caption={Path Integral (\copyright 2018 Shikhar Mittal. All rights reserved).},captionpos=t]
N_sweep=20000;
ep=input('Lattice spacing: ');
h=input('Enter the h for [-h,h] ');
L=input('The anharmonicity parameter: ');
N_t=250/ep;

X2=zeros(1,N_sweep); %To store <x^2> from each path
X4=zeros(1,N_sweep); %To store <x^4> from each path
u=zeros(1,N_t+1);    %Initialise a cold start
U=[];

for j=1:N_sweep
    %accrate=0;     %Uncomment for acceptance rate
    order=randi([2,N_t],1,N_t-1);%Site visiting order

    for i=1:N_t-1   %This loop consitutes 1 sweep
        k=order(i);
        del=2*h*(rand-0.5);
        uc=u(k)+del;
        DS=ep*del*((1+ep^2/2)*(uc+u(k))-(u(k-1)+u(k+1)))+...
            L*ep^5*(uc^4-u(k)^4);

        if(rand<exp(-DS))
            u(k)=uc;
            %accrate=accrate+1/(N_t-1);
        end
    end
    X2(j)=mean(u.^2);
    X4(j)=mean(u.^4);
    if mod(j,100)==0 %100 sweeps skipped for ep>0.2 otherwise use 2*tauint
        U=[U ep*u];
    end
end
[UWX2,UerrX2,ererX2,tauX2,dtauX2]=UWerr(X2',[],[],0,1);
[UWX4,UerrX4,ererX4,tauX4,dtauX4]=UWerr(X4',[],[],0,1);

H=histogram(U,50,'Normalization','pdf');  %Path histogram
fprintf('Bin width of histogram %f\n', H.BinWidth);
xlabel('$x$','Interpreter','latex','Fontsize',20)
ylabel('$|\psi_0|^2$','Interpreter','latex','Fontsize',20)

E_0=ep^2*UWX2+3*L*ep^4*UWX4;    %Virial theorem
ErrE_0=ep^2*UerrX2+3*L*ep^4*UerrX4; %Associated error
fprintf('\n E_0 = %f\n', E_0);
fprintf('Error in E_0 = %f\n', ErrE_0);
\end{lstlisting}
\vspace*{2\baselineskip}
\begin{lstlisting}[style= Matlab-editor,caption={Schr\"odinger Equation (\copyright 2018 Shikhar Mittal. All rights reserved).},captionpos=t]
function SE
global n a
a=input('Anharmonicity = ');
n=input('Enter the n: ');

n1=n;
solinit=bvpinit(linspace(0,10,100),@init,n1);
sol=bvp4c(@sch,@bc,solinit);
u=linspace(0,10,1000); %Solve the BVP on this interval
uo=linspace(-10,10,1999);
Su=deval(sol,u);

SU=Su(1,:);              %Only the wavefunction is needed!
pks=length(abs(findpeaks(SU( abs(SU.^2)>0.01 ).^2)));
if SU(1)==1
    count=2*pks+1;
elseif SU(1)==0
    count=2*pks;
end

for q=1:10
    if count~=n+1
        n1=(n1+sol.parameters);
        solinit=bvpinit(linspace(0,10,100),@init,n1);
        sol=bvp4c(@sch,@bc,solinit);
        Su=deval(sol,u);
        SU=Su(1,:);
        pks=length(abs(findpeaks(SU(abs(SU.^2)>0.01).^2)));
        if SU(1)==1
            count=2*pks+1;
        elseif SU(1)==0
            count=2*pks;
        end
    else
        break;
    end
end

if mod(n,2)==0          %For even function
    psi=[flip(SU(2:end)) SU(1) SU(2:end)]; %Even extension
else                    %For odd function
    psi=[-flip(SU(2:end)) SU(1) SU(2:end)]; %Odd extension
end
N=simp(psi.^2); %Normalisation constant

plot(uo,1/N*psi.^2,'r','LineWidth',1.5)
xlabel('$x$','Interpreter','latex','Fontsize',20)
ylabel('$|\psi_0|^2$','Interpreter','latex','fontsize',20)

u2=simp(uo.^2.*psi.^2)/N; u4=simp(uo.^4.*psi.^2)/N;
fprintf('<u^2>= %0.4f\n',u2)
fprintf('<u^4>= %0.4f\n',u4)
fprintf('\nThe eigen-energy is %0.6f\n',sol.parameters);

%---------------------------------------------------------
function eqns=sch(x,y,E)
global a
eqns=[y(2),(x^2+2*a*x^4-2*E)*y(1)];
%SE is 2nd order; convert it into 2 Ist order eq.

%---------------------------------------------------------
function res = bc(ya,yb,E)
global n
if mod(n,2)~=0
    res=[ya(1),yb(1),ya(2)-1];
else
    res=[ya(1)-1,yb(1),ya(2)];
end
%---------------------------------------------------------
%This function creates an initial guess for the function.
%An obvious choice for the guess would be the known
%solutions of harmonic oscillator.
function psi = init(x)
global n
if n~=0
    psi=[hermiteH(n,x)*exp(-x^2/2),(2*n*hermiteH(n-1,x)...
        -x*hermiteH(n,x))*exp(-x^2/2)];
else
    psi=[hermiteH(0,x)*exp(-x^2/2),-x*hermiteH(0,x)*...
        exp(-x^2/2)];
end
%psi(1)= wavefunction, psi(2) = its derivative
%---------------------------------------------------------
%Simpson's 1/3 rule for numerical integration
function I = simp(y)
I=1/(300)*(y(1)+2*sum(y(3:2:1997))+4*sum(y(2:2:1998))...
    +y(end));
\end{lstlisting}

\end{document}